\documentclass[12pt,preprint]{aastex}

\newcommand{\R}{\theta}
\newcommand{\Pratio}{{\cal R}}
\newcommand{\asinh}{\hbox {asinh}\,}
\newcommand{\RP}{{\R_P}}
\newcommand{\PFluxFactor}{f_2}
\newcommand{\RPcrit}{f_1}
\newcommand{\SBmin}{\mu_{\rm min}}
\newcommand{\RPsmall}{\RP_{\rm min}}
\newcommand{\muh}{{\mu_{50}}}
\newcommand{\Rh}{{\R_{50}}}
\newcommand{\rpsf}{{r_{\rm PSF}}}
\newcommand{\rmodel}{{r_{\rm model}}}
\newcommand{\rfiber}{{r_{\rm fiber}}}
\newcommand{\sbunits}{{{\rm mag}\;{\rm arcsec}^{-2}}}
\newcommand{\ie}{i.e.}

\newcommand{\be}{\begin{equation}}
\newcommand{\ee}{\end{equation}}

\newcommand{\photo}{{\tt photo}}
\newcommand{\Photo}{{\tt Photo}}
\newcommand{\etal}{{et al.}}
\newcommand{\gtapprox}{\mathrel{
   \rlap{\raise 0.511ex \hbox{$>$}}{\lower 0.511ex \hbox{$\sim$}}}}
\newcommand{\ltapprox}{\mathrel{
   \rlap{\raise 0.511ex \hbox{$<$}}{\lower 0.511ex \hbox{$\sim$}}}}

\begin{document}

\title{Spectroscopic Target Selection in the Sloan Digital Sky Survey: The Main Galaxy Sample}
\author{
Michael A. Strauss\altaffilmark{\ref{Princeton}}, 
David H. Weinberg\altaffilmark{\ref{Ohio},\ref{IAS}},
Robert H. Lupton\altaffilmark{\ref{Princeton}}, 
Vijay K.  Narayanan\altaffilmark{\ref{Princeton}},
James Annis\altaffilmark{\ref{Fermilab}},
Mariangela Bernardi\altaffilmark{\ref{Chicago}},
Michael Blanton\altaffilmark{\ref{Fermilab}},
Scott Burles\altaffilmark{\ref{Chicago}},
A. J. Connolly\altaffilmark{\ref{Pittsburgh}},
Julianne Dalcanton\altaffilmark{\ref{Washington}},
Mamoru Doi\altaffilmark{\ref{IoATokyo},\ref{RESCEU}},
Daniel Eisenstein\altaffilmark{\ref{Chicago},\ref{Steward},\ref{Hubble}},
Joshua A. Frieman\altaffilmark{\ref{Fermilab},\ref{Chicago}},
Masataka Fukugita\altaffilmark{\ref{CosmicRay},\ref{IAS}},
James E. Gunn\altaffilmark{\ref{Princeton}}, 
\v{Z}eljko Ivezi\'{c}\altaffilmark{\ref{Princeton}},
Stephen Kent\altaffilmark{\ref{Fermilab}},
Rita S.J. Kim\altaffilmark{\ref{Princeton},\ref{JHU}},
G. R. Knapp\altaffilmark{\ref{Princeton}},
Richard G. Kron\altaffilmark{\ref{Chicago},\ref{Fermilab}},
Jeffrey A. Munn\altaffilmark{\ref{Flagstaff}},
Heidi Jo Newberg\altaffilmark{\ref{Fermilab},\ref{RPI}},
R. C. Nichol\altaffilmark{\ref{CMU}},
Sadanori Okamura\altaffilmark{\ref{DoATokyo},\ref{RESCEU}},
Thomas R. Quinn\altaffilmark{\ref{Washington}},
Michael W. Richmond\altaffilmark{\ref{Rochester}},
David J. Schlegel\altaffilmark{\ref{Princeton}},
Kazuhiro Shimasaku\altaffilmark{\ref{DoATokyo},\ref{RESCEU}},
Mark SubbaRao\altaffilmark{\ref{Chicago}},
Alexander S. Szalay\altaffilmark{\ref{JHU}},
Dan VandenBerk\altaffilmark{\ref{Fermilab}},
Michael S. Vogeley\altaffilmark{\ref{Drexel}},
Brian Yanny\altaffilmark{\ref{Fermilab}},  
Naoki Yasuda\altaffilmark{\ref{NAOJapan}}, 
Donald G. York\altaffilmark{\ref{Chicago}}, and
Idit Zehavi\altaffilmark{\ref{Fermilab},\ref{Chicago}}
}
\newcounter{address}
\addtocounter{address}{1}
\altaffiltext{1}{Princeton University Observatory, Princeton, NJ 08544
\label{Princeton}}
\addtocounter{address}{1}
\altaffiltext{2}{Ohio State University, Dept.~of Astronomy, 140
W. 18th Ave., Columbus, OH 43210
\label{Ohio}}
\addtocounter{address}{1}
\altaffiltext{3}{Institute for Advanced Study, Olden Lane,
Princeton, NJ 08540
\label{IAS}}
\addtocounter{address}{1}
\altaffiltext{4}{Fermi National Accelerator Laboratory, P.O. Box 500,
Batavia, IL 60510
\label{Fermilab}}
\addtocounter{address}{1}
\altaffiltext{5}{The University of Chicago, Astronomy \& Astrophysics
Center, 5640 S. Ellis Ave., Chicago, IL 60637
\label{Chicago}}
\addtocounter{address}{1}
\altaffiltext{6}{Department of Physics and Astronomy,
          University of Pittsburgh,
          Pittsburgh, PA 15260
\label{Pittsburgh}}
\addtocounter{address}{1}
\altaffiltext{7}{University of Washington, Department of Astronomy,
Box 351580, Seattle, WA 98195
\label{Washington}}
\addtocounter{address}{1}
\altaffiltext{8}{Institute of Astronomy, School of Science, University of Tokyo,
Mitaka, Tokyo, 181-0015 Japan 
\label{IoATokyo}}
\addtocounter{address}{1}
\altaffiltext{9}{Research Center for the Early Universe, 
School of Science, University of Tokyo, Tokyo, 181-0033 Japan
\label{RESCEU}}
\altaffiltext{10}{Steward Observatory, University of Arizona, 933
N. Cherry Ave., Tucson, AZ 85721
\label{Steward}}
\addtocounter{address}{1}
\altaffiltext{11}{Institute for Cosmic Ray Research, University of
Tokyo,  Kashiwa, Chiba, 277-8582 Japan  
\label{CosmicRay}}
\addtocounter{address}{1}
\altaffiltext{12}{
Department of Physics and Astronomy, The Johns Hopkins University,
   3701 San Martin Drive, Baltimore, MD 21218, USA
\label{JHU}
}
\addtocounter{address}{1}
\altaffiltext{13}{U.S. Naval Observatory, Flagstaff Station, 
P.O. Box 1149, 
Flagstaff, AZ  86002-1149
\label{Flagstaff}}
\addtocounter{address}{1}
\altaffiltext{14}{Physics Department, Rensselaer Polytechnic
  Institute, SC1C25, Troy, NY 12180
\label{RPI}}
\addtocounter{address}{1}
\altaffiltext{15}{Dept. of Physics, Carnegie Mellon University,
     5000 Forbes Ave., Pittsburgh, PA-15232
\label{CMU}}
\addtocounter{address}{1}
\altaffiltext{16}{Department of Astronomy, School of Science, University of Tokyo,
Tokyo, 181-0033 Japan 
\label{DoATokyo}}
\addtocounter{address}{1}
\altaffiltext{17} {Physics Department,
            Rochester Institute of Technology,
            85 Lomb Memorial Drive,
            Rochester, NY 14623-5603
\label{Rochester}}
\addtocounter{address}{1}
\altaffiltext{18}{Department of Physics, Drexel University,
  3141 Chestnut St., Philadelphia, PA 19104
\label{Drexel}}
\addtocounter{address}{1}
\altaffiltext{19}{National Astronomical Observatory, 
Mitaka, Tokyo 181-8588, Japan
\label{NAOJapan}}
\addtocounter{address}{1}
\altaffiltext{20}{Hubble Fellow
\label{Hubble}}

\begin{abstract}

We describe the algorithm that selects the main sample of galaxies for
spectroscopy in the Sloan Digital Sky Survey from the photometric data
obtained by the imaging survey.  Galaxy photometric properties are
measured using the Petrosian magnitude system, which measures flux in
apertures determined by the shape of the surface brightness profile.
The metric aperture used is essentially independent of cosmological
surface brightness dimming, foreground extinction, sky brightness, and
the galaxy central surface brightness.  The main galaxy sample
consists of galaxies with $r$-band Petrosian magnitude $r \leq 17.77$
and $r$-band Petrosian half-light surface brightness $\muh \leq 24.5$
magnitudes per square arcsec.  These cuts select about 90 galaxy
targets per square degree, with a median redshift of 0.104.  We carry
out a number of tests to show that (a) our star-galaxy separation
criterion is effective at eliminating nearly all stellar contamination
while removing almost no genuine galaxies, (b) the fraction of
galaxies eliminated by our surface brightness cut is very small ($\sim
0.1\%)$, (c) the completeness of the sample is high, exceeding 99\%,
and (d) the reproducibility of target selection based on repeated
imaging scans is consistent with the expected random photometric
errors.  The main cause of incompleteness is blending with saturated
stars, which becomes more significant for brighter, larger galaxies.
The SDSS spectra are of high enough signal-to-noise ratio ($S/N > 4$
per pixel) that
essentially all targeted galaxies $(99.9\%)$ yield a reliable
redshift (i.e., with statistical error $< 30\rm \, km\,s^{-1}$).
About 6\% of galaxies that satisfy the selection criteria 
are not observed because they have a companion closer than the $55''$
minimum separation of spectroscopic fibers, but these galaxies can be accounted
for in statistical analyses of clustering or galaxy properties.  The
uniformity and completeness of the galaxy sample make it ideal for
studies of large scale structure and the characteristics of the galaxy
population in the local universe.

\end{abstract}
\keywords{surveys --- galaxies:distances and redshifts --- galaxies:photometry}

\section{Introduction}

The Sloan Digital Sky Survey (SDSS; \citealt{York}) is carrying out an
imaging survey in five photometric bands of $\pi$ ster in the
north Galactic cap, and a follow-up spectroscopic survey of
roughly $10^{6}$ galaxies and $10^{5}$ quasars, complete within precisely
defined selection criteria.   
The main scientific drivers of the SDSS
are the
large-scale distributions of galaxies and quasars.  In order to carry
out precise measurements of galaxy clustering on the
largest scales, and to measure the distribution of galaxy properties with the
highest possible precision, it is necessary that the sample of
galaxies for which spectra are taken be selected in a uniform and
objective manner.  
The northern spectroscopic survey targets two samples
of galaxies: a flux-limited sample
to $r=17.77$ (hereafter called the main sample) 
and a flux- and color-selected sample extending to $r=19.5$, designed to 
target luminous red galaxies (LRGs).   This paper describes the
algorithm used to
select the main galaxy sample and presents demonstrations
that the algorithm meets the survey goals of uniformity and
completeness.  A separate paper (\citealt{LRG}) discusses the LRG sample.  

\subsection {The Sloan Digital Sky Survey}
\label{sec:SDSS}

The SDSS hardware, software, and data products 
are summarized by \citet{York} and \citet{EDR}. 
In brief, the survey is carried out using a dedicated, wide-field
2.5m telescope,
a mosaic CCD camera (\citealt{Gunn}), 
two fiber-fed double spectrographs, and an
auxiliary 0.5 m telescope for photometric calibration. 
The imaging is done in drift scan mode
with the 30 photometric CCDs of
the mosaic camera imaging $\approx 20$ square degrees per hour,
in five broad bands, $u,g,r,i$ and $z$ \citep{Fukugita:1996} that cover the
entire optical range from the atmospheric ultraviolet cutoff in the
blue to the sensitivity limit of silicon CCDs in the red.
The imaging data are 95\% complete for point sources at
$r^* \approx 22.2$, and the
photometric calibration is accurate to 
about 3\% in $r$ at this writing (\citealt{Hogg:2001,Smith:2002}).  
Because this calibration is still preliminary, we will refer to 
current measurements with the notation $u^*$, $g^*$, $r^*$,
$i^*$, $z^*$, but we use $u$, $g$, $r$, $i$, $z$ to refer to
the SDSS filter and magnitude system itself.\footnote{This notation
is a change from some earlier papers, including \citet{Fukugita:1996},
which referred to the filter system as $u^\prime$, $g^\prime$,
$r^\prime$, $i^\prime$, $z^\prime$; see the discussion by \citet{EDR}.}
The astrometric calibration \citep{Pier:2002} is done
by comparison with the Tycho-2 \citep{hoeg00} and UCAC
\citep{zacharias00} standards, and is accurate to  
0.1 arcsec rms per coordinate.

  The imaging data are reduced using a series of interlocking
pipelines (\photo; \citealt{adass}),
which flat-field the data, find all objects, match up
detections in the different bands and perform measurements of their
properties, and apply the photometric and astrometric calibrations.
Spectroscopic targets --- 
the galaxies described in this paper, LRGs (\citealt{LRG}),
quasars (\citealt{QSO}), and a variety of other categories of objects \citep{EDR},
are chosen from the resulting catalog of detected objects.  

The spectroscopic component of the survey is carried out using two
fiber-fed double spectrographs, covering the wavelength range 3800\AA\
to 9200\AA\ over 4098 pixels.  They have a resolution 
$\lambda/\Delta\lambda$ varying between
1850 and 2200, and together they are fed by 640 fibers, each with an
entrance diameter of $3''$.  The fibers are manually plugged into
plates inserted into the focal plane; the mapping of fibers to plates
is carried out by a tiling algorithm (\citealt{tiling}) that optimizes
observing efficiency in the presence of large-scale structure.  The
finite diameter of the fiber cladding prevents fibers on any given
plate from being placed closer than $55''$ apart.

For any given plate, a series of fifteen-minute exposures is carried
out until the mean signal to noise ratio (S/N) per resolution element
exceeds 4 for objects with fiber magnitudes (i.e., as measured through
the $3''$ aperture of the fiber) brighter than $g^*=20.2$
and $i^*=19.9$,
as determined by preliminary reductions done at the observing site.
Under good conditions (dark, clear skies and good seeing), this
typically requires a total of 45 minutes of exposure.

\subsection{The Main Galaxy Spectroscopic Sample}

The main galaxy spectroscopic survey is fully sampled to its magnitude
limit within the survey footprint, which is planned to be an
elliptical area of extent $110^\circ\times 130^\circ$, chosen to
minimize Galactic extinction and maximize observing efficiency.  
The median redshift of this sample is $z \approx 0.1$.  This
large galaxy sample will allow us to measure many independent modes of
the density fluctuations on scales comparable to the peak of the galaxy
 power spectrum, largely free from the aliasing that can affect
surveys with at least one narrow dimension (cf., Kaiser \& Peacock
1991; Tegmark 1995).  For some instrumental set-ups and scientific
goals (e.g., low-order measures of large scale clustering), one can
gain efficiency by sparse sampling, \ie, by observing only a fraction
of the galaxies down to some limiting magnitude (cf., Kaiser 1986).
However, sparse sampling adversely affects other kinds of
investigation, including group and cluster studies, high-order
clustering measures, and recovery of the underlying galaxy density
field (see, e.g., \citealt{SS96}).  Moreover, the field of view and
number of spectroscopic fibers of the SDSS were chosen to allow simultaneous
spectroscopy of essentially {\em all} the galaxies in a given field to the
faintest magnitude for which the 2.5m telescope can measure redshifts
in a reasonable amount of time.  We have therefore opted for complete
sampling in the main galaxy redshift survey.  

We wish to select a magnitude-limited galaxy sample.   We have carried out
the selection in a single observed band for simplicity. 
We wish the galaxy detection and photometric measurement in that
band to be of high S/N, and we prefer a
red passband so that $K$ corrections are modest, fluxes are determined
mainly by the older stars that dominate the stellar mass, and 
uncertainties in Galactic reddening make little difference to
the inferred galaxy magnitude.
In the SDSS filter system, this implies either the $r$ or $i$ band. 
We adopt the former because the sky background is
brighter and more variable in the $i$ band than in the $r$ band.
The use of a red bandpass tilts the sample slightly towards galaxies
of earlier morphological type, but at these bright magnitudes, the
$g-r$ color distribution of galaxies is quite narrow \citep{Ivezic02},
and the distribution of galaxy types 
is not radically different from what we would obtain with $g$-band
selection (see also Fig.~4 of \citealt{Shimasaku01}).

  Although we will detail a number of subtleties below, the basic
procedure that we use to define galaxy magnitudes and select
spectroscopic targets can be summarized as follows. Star-galaxy
separation is carried out by comparing the exponential or de Vaucouleurs
model magnitude of an object to its Point Spread Function (PSF)
magnitude.  We define the (angular) Petrosian radius $\RP$ of a galaxy
to be the radius at which the local surface brightness in an annulus
about $\RP$ is 1/5 of the mean surface brightness within $\RP$.  We
define the $r$-band Petrosian magnitude of a galaxy, $r_P$, based on
the flux within a circular aperture of radius $2\,\RP$.  In the
absence of seeing effects, the Petrosian magnitude measures the light
within a well-defined metric aperture on any given galaxy which is
independent of its redshift or foreground extinction.  We define the
half-light surface brightness $\muh$ to be the mean surface brightness
within a circular aperture containing half of the Petrosian flux.  The
main galaxy sample consists of galaxies with $r_P \leq 17.77$ and
$\muh \leq 24.5$ magnitudes per square arcsec, after correcting for
Galactic extinction following Schlegel, Finkbeiner, \& Davis (1998;
hereafter SFD).

The outline of this paper is as follows.  In \S~\ref{sec:properties},
we describe our goals for the target selection algorithm.  
We 
discuss the measurements of Petrosian quantities in detail in 
\S~\ref{sec:petrosian} and Appendix~\ref{sec:petro-appendix}.  The target
selection algorithm itself is described in \S~\ref{sec:algorithm}.
Various tests to show that the algorithm meets the survey requirements are
described in \S~\ref{sec:test}.  
We conclude in
\S~\ref{sec:conclusions}. 

\section{Desired properties of the Selection Algorithm}
\label{sec:properties}

The SDSS spectroscopic galaxy surface density is roughly 100 galaxies
per square degree.  From studies of galaxy number counts
(\citealt{Yasuda}), a fully sampled magnitude-limited survey reaches
this surface density at a magnitude limit of $r \approx 18$.  There
are several desiderata for a galaxy spectroscopic target selection
algorithm:

\begin{enumerate}
\item The selection algorithm should allow accurate determination of a
selection function, whereby the probability that a galaxy with given
properties (magnitude, color, surface brightness, redshift, position
on the sky, presence of neighbors) is targeted can be objectively quantified.  To the extent
possible, this selection function should depend only on redshift, and
should be independent of seeing, stellar contamination, Galactic
extinction, and spectroscopic observing conditions.

\item The algorithm should be based on physically meaningful
parameters which are tied as closely
as possible to the properties of the galaxies.  We must be able to
measure these parameters accurately for the sample of galaxies whose
spectra we plan to obtain.

\item The algorithm should select a uniform sample of 
galaxies with a wide range of physical properties, without biasing against,
for example, galaxies of unusual color or low surface brightness. 

\item The algorithm should select galaxies for which we are
able to obtain a spectrum of sufficient quality to yield a redshift in
the nominal exposure time.  One area of concern in this context will
be galaxies with low surface brightness, for which the total light
down the $3''$ entrance aperture of the fiber will be small. 

\item Finally, the selection algorithm should be simple,
and its behavior should be straightforward and easy to understand.
This makes it easier to test the algorithm,
and it will facilitate the construction of realistic 
mock catalogs of the SDSS galaxy redshift survey from numerical 
simulations of large-scale structure (e.g., \citealt{cole98,colley00}).
\end{enumerate}

  These desiderata do not always point in the same direction.  For
example, one might maximize the redshift success by selecting on
the $3''$ fiber magnitude.  However, this measures a fraction of the
galaxy light that is strongly dependent on redshift and the
atmospheric seeing during the imaging observations, making it quite difficult
to determine a meaningful selection function; moreover, it biases
strongly against low-surface brightness galaxies.  Similarly,
isophotal magnitudes measure a fraction
of the total galaxy light that depends on foreground extinction, sky
brightness (if the isophotal threshold is set relative to the sky
level), and redshift (because of cosmological surface brightness dimming).  
After weighing a number of options, we have settled on an algorithm
that employs a modified form of the \citet{Petrosian} magnitude system,
with galaxies selected in $r$-band.  

\section{Petrosian Quantities}
\label{sec:petrosian}

\subsection{Definition of Petrosian Quantities}

The \citet{Petrosian} magnitude is based on the flux within an
aperture defined by the ratio of the local surface brightness to
the mean interior surface brightness.  
The size of this aperture depends on the {\it shape} of the galaxy's
radial surface brightness profile but not its amplitude.

Let $I(\R)$ be the azimuthally averaged surface
brightness profile of a galaxy, as a function of angular distance from
its center, $\R$.  We define the {\em Petrosian ratio} as the ratio of
the surface brightness in an annulus $0.8\R-1.25\R$ to the mean
surface brightness within $\R$,
\begin{equation} 
\Pratio(\R) = \frac{2 \pi \int_{0.8 \R}^{1.25 \R} I(\R') \R' d\R' / 
  \left[ \pi ((1.25 \R)^2 - (0.8 \R)^2)\right]}
  {2 \pi \int_0^{\R} I(\R') \R' d\R' / (\pi \R^2)}\ .
\label{eq:RPetro}
\end{equation}
The use of a fairly thick annulus reduces the sensitivity
of $\Pratio(\R)$ to noise and to small scale fluctuations in $I(\R)$.
We define the Petrosian radius $\RP$ by the implicit equation
\begin{equation} 
\Pratio(\RP) = \RPcrit\ ,
\label{eq:petrorad}
\end{equation}
where $\RPcrit$ is a constant, which we set to 0.2.
The Petrosian {\em flux} is defined as the flux within a circular aperture
of radius $\PFluxFactor$ times the Petrosian radius,
\begin{equation} 
F_P = 2 \pi \int_0^{\PFluxFactor \RP} I(\R') \R' d\R'\ ,
\label{eq:petroflux}
\end{equation}
and we set $\PFluxFactor = 2$ (hereafter, we refer to 
$\PFluxFactor \RP$ as the {\em Petrosian aperture}).  It will also be
useful to formally define a total flux, $F_{\rm tot}$, as the result of
the integral in equation~(\ref{eq:petroflux}) out to infinity. 
The choice of $f_1$ and $f_2$ is discussed in \S\ref{sec:f1_f2} below.
Note that equation~(\ref{eq:petrorad}) for the Petrosian radius
may have more than one solution; 
in this case, we take the outermost of the solutions.  The technical
details of how \photo\ measures all these quantities are given in 
Appendix~\ref{sec:petro-appendix}.  Note that the images of
overlapping galaxies are deblended using a robust code that conserves
flux (Lupton \etal, in preparation). 

The Petrosian magnitude is defined from the Petrosian flux in the
usual way, once the conversion between detected counts and calibrated
fluxes is determined.  Note, however, that this conversion for SDSS
photometry is strictly valid only for point sources, for which a
proper aperture correction can be quantified; we ignore this
complication in what follows.
Our magnitude system is based on the AB$_{95}$ system
(\citealt{Fukugita:1996}); i.e., 
the mean flux density over any of the broad pass-bands is
$\overline{f}=3631 \times 10^{-0.4m}$ Jy,
to a fair approximation (see the caveats in \citealt{EDR}).
We use the asinh magnitude definition of \citet{asinhMags}, which
behaves well in the regime of low S/N, though for the bright
galaxies in the spectroscopic sample the difference between asinh
magnitudes and traditional logarithmic magnitudes is negligible.
We refer to the $r$-band Petrosian apparent magnitude as $r_P$
and to PSF and model magnitudes (see \S\ref{sec:star-galaxy} below)
as $\rpsf$ and $\rmodel$.

Our target selection algorithm also requires a measure of surface
brightness, for which we want to retain the desirable properties
of the Petrosian system.  The surface brightness within $\RP$ is
an obvious choice, but it turns out to be rather noisy.
Instead, we define the Petrosian 
half-light radius $\Rh$ as that which encloses half
the Petrosian flux,
\begin{equation} 
\int_0^{\Rh} I(\R') \R' d\R' = 0.5\int_0^{\PFluxFactor \RP} I(\R') \R' d\R' \ ,
\label{eq:petroR50}
\end{equation}
and use the mean surface brightness within this radius,
\begin{equation} 
\muh = r_P + 2.5 \log\left[ 2 \,\pi \Rh^2\right].
\label{eq:petroSB}
\end{equation}
Because the flux within $2\RP$ is insensitive to small errors
in $\RP$, the quantities $\Rh$ and $\muh$ can be robustly measured.

\begin{figure}
\plotone{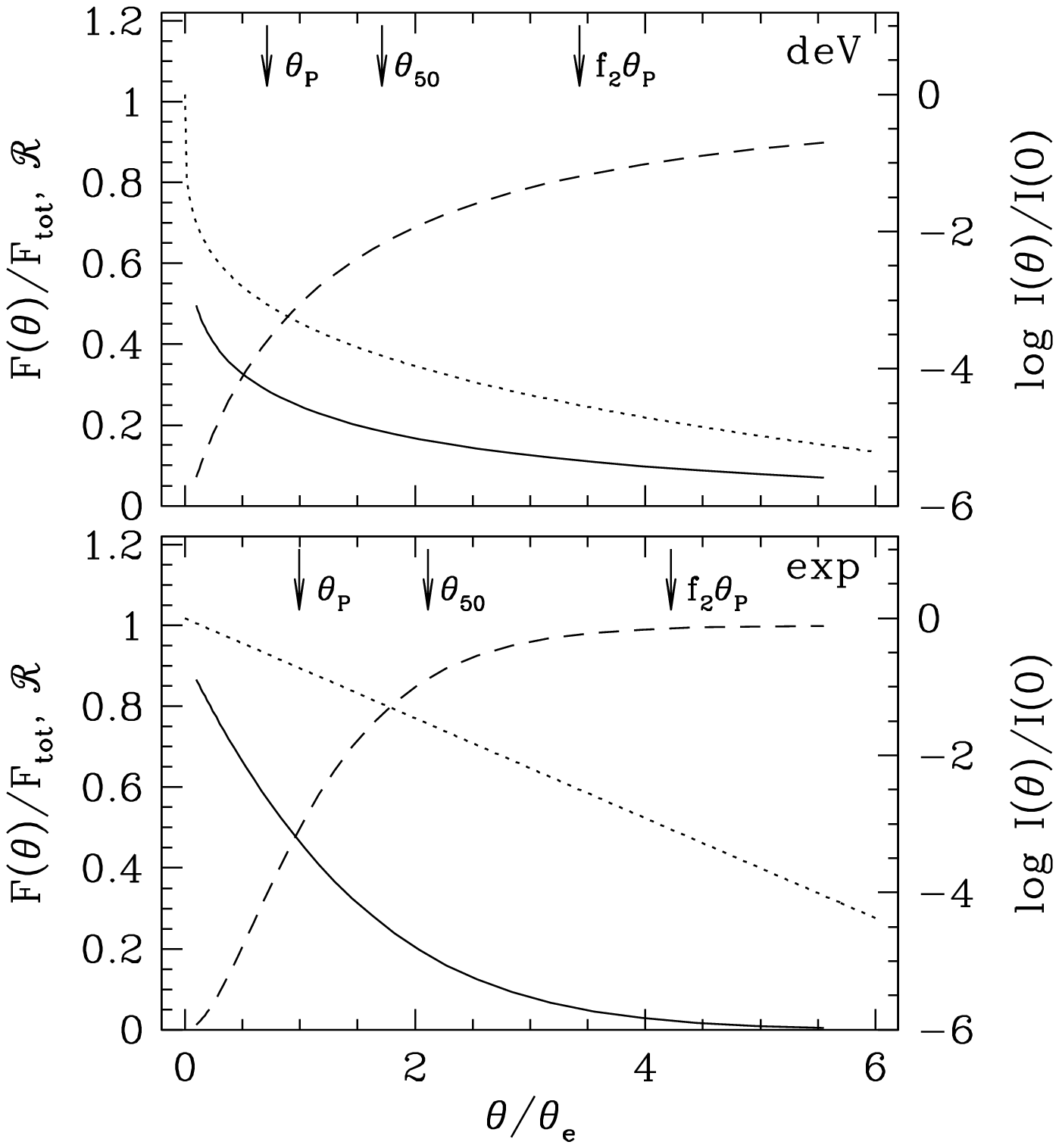}
\caption{Illustration of the Petrosian aperture procedure for a
de Vaucouleurs profile (top) and an exponential profile (bottom),
assuming an axis ratio of one and negligible seeing.
In each panel, the dashed curve shows the curve of growth
(fraction of total light within radius $\R$), and the solid 
curve shows the Petrosian ratio $\Pratio(\R)$.  
The dotted curve shows the logarithmic surface brightness profile,
using the right-hand axis scale.  The central arrow marks
the Petrosian radius at $\Pratio(\R)=0.2$.  Outer and inner
arrows represent the radius of the Petrosian aperture 
($\PFluxFactor \RP$) and the Petrosian half-light radius $\Rh$,
respectively.  All radii are scaled to the true half-light radius
$\R_e$, which is 1.678 scale lengths for the exponential
profile.
}
\label{fig:sbprof}
\end{figure}

Figure~\ref{fig:sbprof} illustrates our definitions for the case of a circular de Vaucouleurs profile
(top; $I(\R) = I_0 \exp[-7.67(\R/\R_e)^{-/4}]$) and a
face-on exponential disk (bottom; $I(\R) = I_0 \exp[-1.68\,\R/\R_e]$).
Dotted, dashed, and solid curves show the surface brightness 
profile, curve of growth, and Petrosian ratio $\Pratio(\R)$.
Arrows mark the Petrosian half-light radius $\Rh$, the Petrosian
radius $\RP$ at which 
$\Pratio=\RPcrit=0.2$ and the Petrosian aperture at
$\PFluxFactor\RP=2\RP$. The Petrosian radius corresponds to 2.1
effective (or half-light) radii (3.5 scale lengths)
for an exponential profile and 1.7 effective radii for a
de Vaucouleurs profile.  
The $2\RP$ Petrosian aperture encompasses
99\% and 82\% of the galaxy's total light in the two cases.
The Petrosian half-light radius $\Rh$ is slightly smaller than the true
half-light radius,
since the Petrosian flux is less than the total flux.

\begin{figure}
\plotone{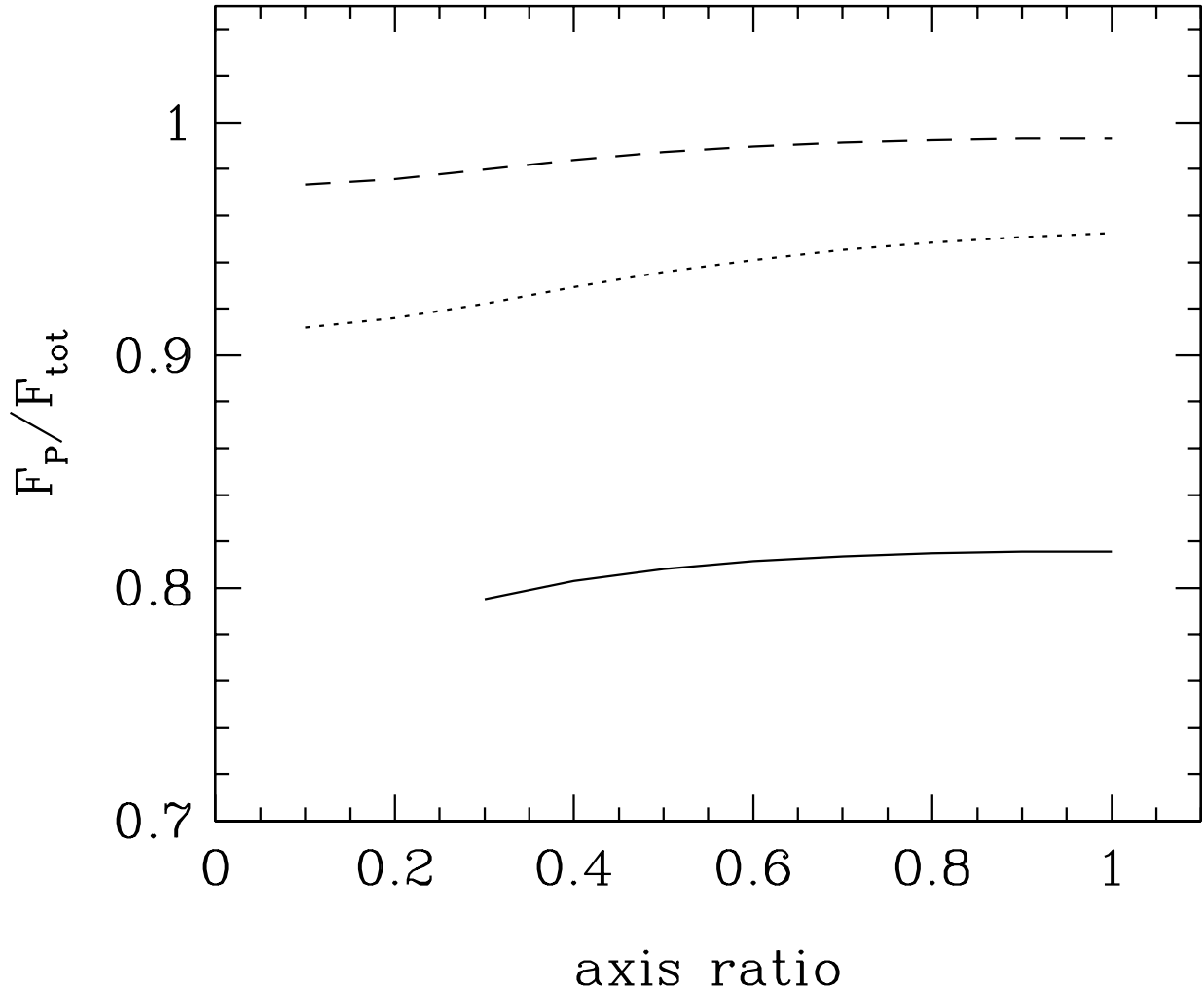}
\caption{Effect of axis ratio on the Petrosian flux, measured using
our circular aperture definitions of Petrosian quantities.
The dashed line shows the fraction of the total flux within the
Petrosian aperture for inclined exponential disks as a function
of axis ratio.  The solid line shows the same quantity for a
de Vaucouleurs law galaxy.  The dotted line represents a galaxy
with an inclined (exponential) disk and a circular (de Vaucouleurs
law) bulge, assuming a 1:1 bulge-to-disk ratio and a bulge half-light 
radius that is half that of the disk.
}
\label{fig:inclination}
\end{figure}

Note that we use circular apertures rather than elliptical apertures
for all measurements.  Elliptical apertures are difficult to choose
for galaxies whose light distributions are not well described by
concentric self-similar ellipses.  Moreover, for disk galaxies, the
circular-aperture surface brightness profile is also less sensitive to
inclination than the elliptical-aperture profile, at least to the
extent that internal extinction can be neglected.  Because the
Petrosian aperture is always large enough to contain most of a
galaxy's light, the ratio of the Petrosian flux to total flux is
insensitive to inclination (or de Vaucouleurs axis ratio), as shown in
Figure~\ref{fig:inclination}.

In the absence of noise, the Petrosian aperture is unaffected  by
foreground extinction or by the cosmological dimming of the surface brightness.  
Thus, identical galaxies seen at 
two different (luminosity) distances have fluxes related
exactly as distance$^{-2}$ (in the absence of K corrections).  
One can therefore determine the maximum distance at which a galaxy
would enter a flux-limited sample without knowing the galaxy's
surface brightness profile (which would be needed for the
equivalent calculation with, e.g., isophotal magnitudes).
Moreover, two galaxies that have the same surface brightness
profile shape but different central surface brightness
have the same fraction of their flux represented in the Petrosian
magnitude, so there is no bias against the selection of low surface
brightness galaxies of sufficiently bright Petrosian magnitude.
In the absence of noise, the Petrosian magnitude is independent
of sky brightness, and for the large angular extent of the galaxies of 
the spectroscopic
sample, it is also insensitive to seeing. 

\subsection{Setting $\RPcrit$ and $\PFluxFactor$}
\label{sec:f1_f2}

  Here we describe the rationale behind our choice for the Petrosian
parameters $\RPcrit=0.2$ and $\PFluxFactor=2$.
Setting $\RPcrit$ too high would increase
our sensitivity to seeing (by making $\RP$ small), while
setting it too low would require measuring the 
Petrosian ratio at a point that the surface brightness is many
magnitudes below that of the sky, 
making us particularly sensitive to sky subtraction effects.   
If the Petrosian aperture $\PFluxFactor$ is too large,
the measurement of the Petrosian flux is badly affected by
sky noise and uncertainty in the sky level.  If the aperture is too small, on the other hand,
then the Petrosian magnitude departs substantially from
the theoretical ideal of a total magnitude,
and, equally important, it becomes sensitive to seeing and to noise in the
measurement of $\RP$.

We have carried out extensive simulations of galaxies with realistic
distributions of surface brightness and bulge-to-disk ratio (following
\citealt{Fukugita:1998}),
and have processed the resulting images through
\photo\ with a range of choices for the Petrosian parameters.  
We find that seeing
has an appreciable effect on the Petrosian quantities for galaxies 
that would be in
the spectroscopic sample, when $\RPcrit \ge 0.25$.  
In the simulations, only 2\% of the galaxies fail to have a Petrosian
radius measured (i.e., i.e., the S/N is too low to measure the
Petrosian ratio down to $\RPcrit$) at the spectroscopic limit for
$\RPcrit = 0.2$, but the fraction of such failures becomes appreciable at
smaller $\RPcrit$.  
See \citet{Yagi} for a similar discussion. 
Note that \photo\ still reports a reasonable
measure of a ``Petrosian'' magnitude even if it is unable to measure a
Petrosian radius for a given object (Appendix~\ref{sec:petro-appendix}).

\begin{figure}
\plotone{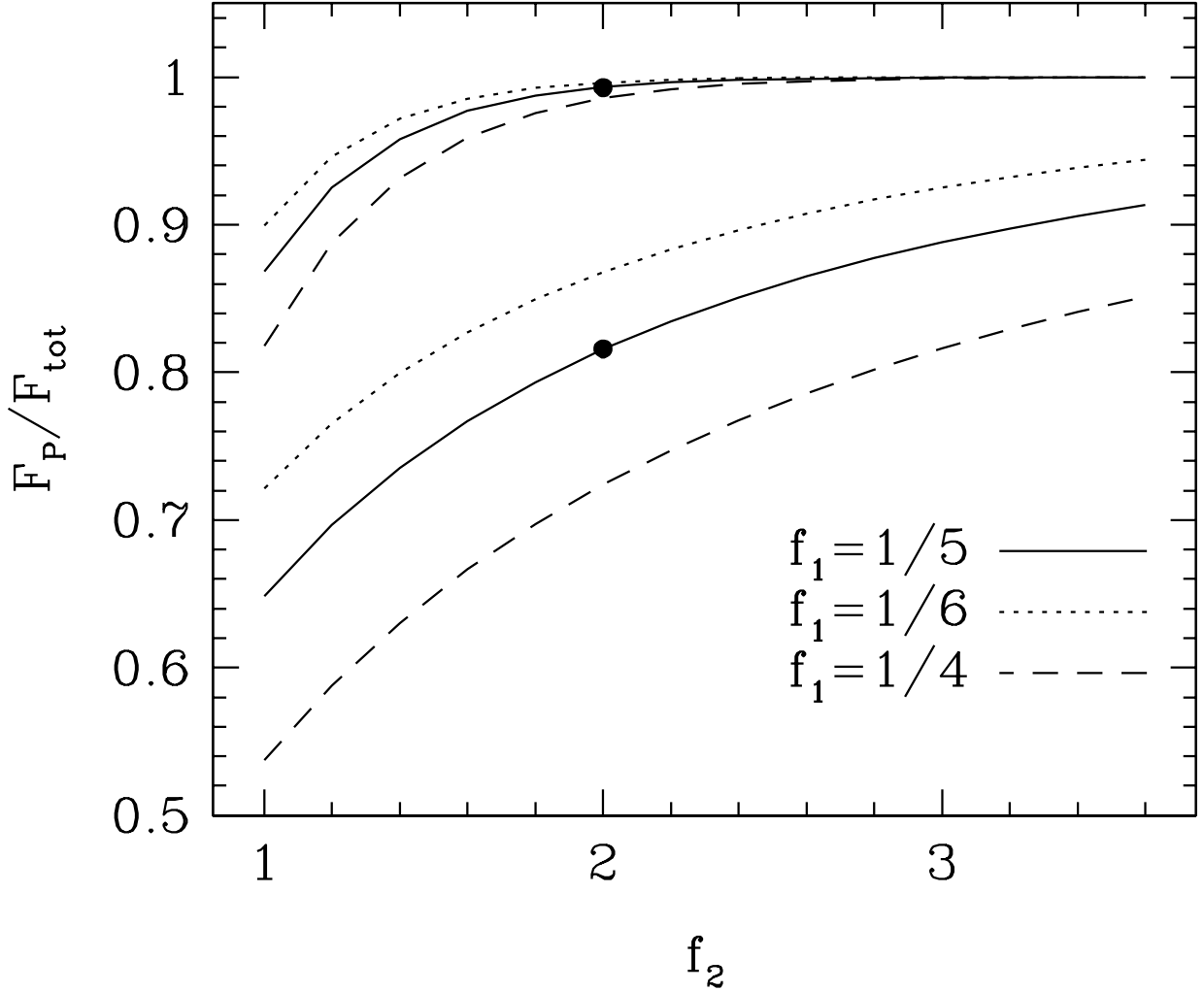}
\caption{
The fraction of the total light within the Petrosian aperture
$\PFluxFactor\RP$ as a function of the value of $\PFluxFactor$,
for $\RPcrit=1/5$ (solid lines), $1/6$ (dotted lines), and $1/4$
(dashed lines).  Lower curves correspond to de Vaucouleurs law
profiles and upper curves to exponential profiles.  Filled circles
mark our adopted values $\RPcrit=1/5$, $\PFluxFactor=2$.
These calculations simulate idealized circular galaxies, and seeing is assumed
to be negligible. 
}
\label{fig:f1f2}
\end{figure}

Given a value of $\RPcrit$, the quantity $\PFluxFactor$ sets the Petrosian
aperture.  Figure~\ref{fig:f1f2} shows the fraction of light
within the Petrosian aperture for various combinations of $f_1$ 
and $f_2$ (the effect of seeing is not included here, but for a given
value of $f_1$, the dependence of $F_P/F_{tot}$ on seeing is quite
weak; see further discussion below).  For many reasonable combinations 
of Petrosian parameters, one gets almost 100\% of 
the light in the Petrosian aperture for galaxies with exponential profiles.  
De Vaucouleurs profiles are much more extended, however, and
an appreciable fraction of their light lies in the regime 
in which the S/N per pixel is appreciably below unity.  
An aperture large enough to enclose, say, 95\% of the light of
a de Vaucouleurs profile includes many low S/N pixels and thus
a substantial amount of sky noise.  For $\RPcrit=0.2$, the
fraction of light included in the Petrosian aperture for a de
Vaucouleurs profile rises from 82\% to 89\% as $\PFluxFactor$
ranges from 2 to 3.  However, the simulations mentioned above show that
the photometric errors in Petrosian magnitudes at the galaxy
spectroscopic survey limit 
increase from 0.03 mag to 0.09 mag over this $\PFluxFactor$ range.
We have therefore settled on the value $\PFluxFactor = 2$,
to keep the S/N of Petrosian magnitudes high while still retaining
most of the light for de Vaucouleurs galaxies.

For any given morphological type (at least as defined by
the surface brightness profile), the correction from a Petrosian to a 
total magnitude is an additive constant (fixed multiplicative
factor for the flux), only weakly dependent upon inclination
(Figure~\ref{fig:inclination}).  Any scientific analysis 
that uses the redshift survey must consider whether a
different fraction of light is included for ellipticals and spirals
affects the result.  This is important for such analyses as the total
luminosity density of the universe (e.g., \citealt{Yasuda,Blanton}), but
it does not enter in calculating the radial selection function for
large-scale structure studies (as the fraction of light is independent
of redshift).  This dependence of the fraction of the light measured on the
galaxy light profile is unavoidable for any target selection algorithm.

\subsection{Effects of seeing, redshift, and sky background}
\label{sec:systematics}

 The effect of seeing on Petrosian quantities is not completely negligible, as is
shown by Figures 1 and 2 of \citet{Blanton}.
For a poorly resolved galaxy, the surface brightness profile
approaches that of a PSF, and the ratio of Petrosian flux to total flux
approaches 0.95, the value for a PSF.  Thus, as the 
seeing radius approaches the half-light radius, the Petrosian
flux of an exponential galaxy is biased downward, and the Petrosian
flux of a de Vaucouleurs law galaxy is biased upward.  In practice,
a galaxy with $\Rh=2''$ observed in $1.5''$ seeing will have its
Petrosian magnitude biased by 1-3\%
depending on profile and axis ratio, and these effects are much smaller
for larger galaxies (see \citealt{Blanton}, Fig. 1).
Roughly 35\% of galaxies in the spectroscopic sample have
$\Rh<2''$. 
At the spectroscopic magnitude limit, the typical measurement error in
Petrosian magnitudes is $\sim 0.035$ magnitudes
(see Figure~\ref{fig:magerr} and the accompanying discussion in
\S\ref{sec:reproducibility} below), so seeing effects are small
compared to photon noise for the great majority of galaxies in 
the spectroscopic sample.

\begin{figure}
\plotone{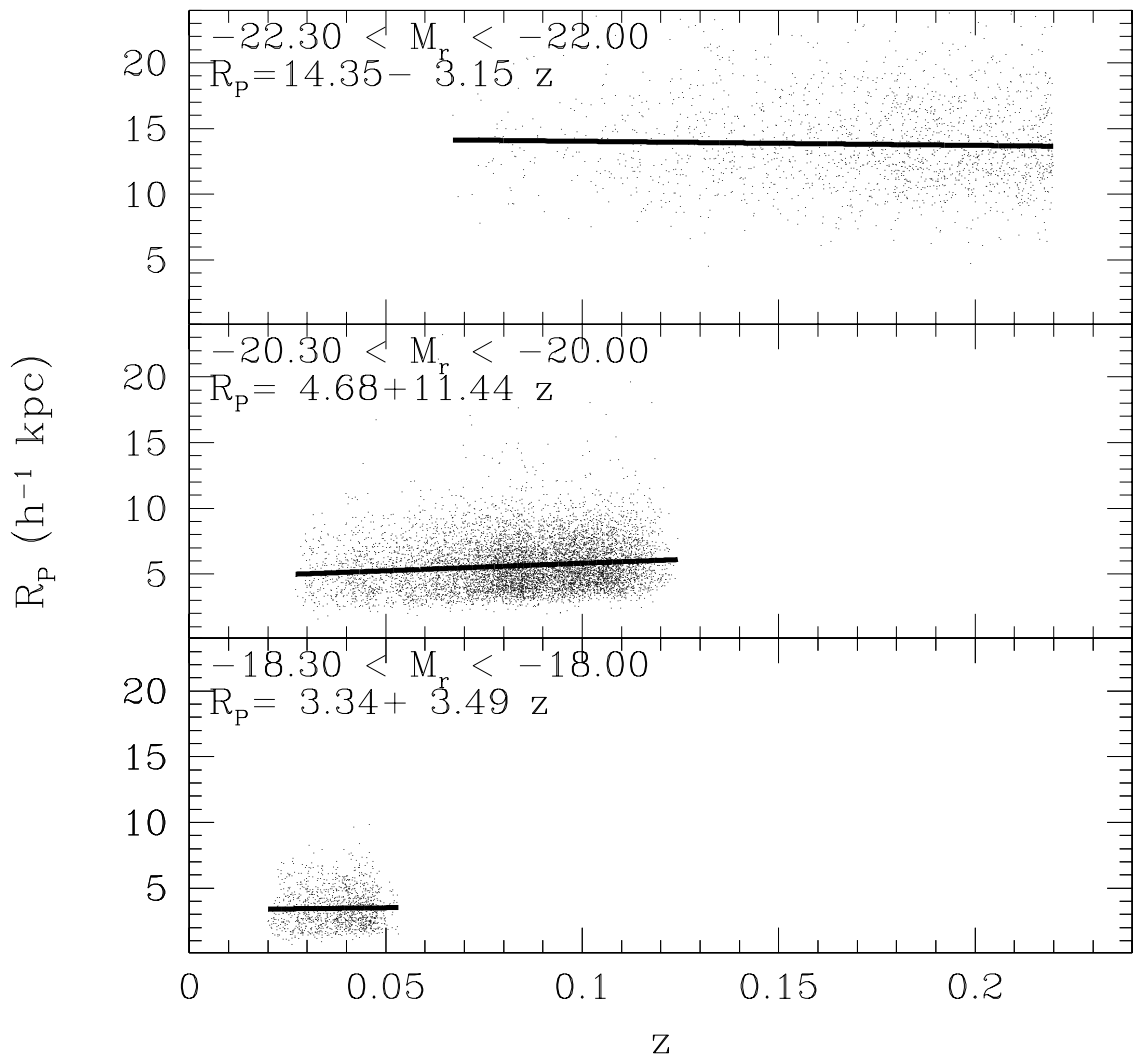}
\caption{$R_P$, the Petrosian radius $\RP$ multiplied by the angular 
diameter distance
(assuming $\Omega_m=0.3$
and $\Omega_\Lambda=0.7$) as a function of redshift $z$ for several
small ranges of absolute magnitude in the main galaxy sample; in each
panel, we show a linear regression of $R_p$ along $z$ as the solid
line.  
} 
\label{fig:redshift-radius}\end{figure}

A related issue is the scaling of Petrosian magnitudes with redshift.
Figure~\ref{fig:redshift-radius} shows the redshift dependence of
the measured $R_p$, the Petrosian radius expressed in $h^{-1}$ kpc, for galaxies in
three narrow slices in absolute magnitude from the SDSS redshift survey.  In the absence of
evolution and seeing effects, there should be no redshift dependence
at all. 
In the middle panel, the redshift dependence causes a 20\% increase in
the measured Petrosian radius over the redshift range spanned; for a
typical galaxy in the SDSS, the last 20\% in radius contains about 5\%
of the galaxy flux, which is thus an upper limit to the expected
systematic effects on redshift (Blanton \etal, in preparation,
conclude that the systematic effects are in fact considerably smaller
than this).

Photometry of objects requires a model for the underlying sky
brightness over the extent of the object.  For objects that have at
least one pixel in any band with flux greater than $200\sigma_{\rm
sky}$, where $\sigma_{\rm sky}$ is the rms amplitude of fluctuations in
sky level in that band (this corresponds roughly to objects brighter
than $r^* = 17.5$), \photo\ measures the magnitudes twice, using two
different models for the sky brightness. 
The first measurement (called the {\tt BRIGHT} measurement)
uses a global sky value determined over an entire frame, and the second uses 
a model for the local sky estimated by
median-smoothing the image on a scale of approximately $100''$, and
thus will be biased high by objects of large angular extent.  
The difference between these two magnitudes of an object is a
reasonable estimate of the photometric 
error arising from uncertainties in determining the sky underneath an
object; in particular, one might be concerned that galaxies of large
angular extent will artificially raise the estimate of the local sky,
therefore biasing their photometry low.  Using 
210 galaxies with $r^* < 16$ from the SDSS Early Data
Release \citep{EDR}, we found that 90\% of the objects lie in the
range 
$-0.020 < r^*({\rm local\ sky}) - r^*({\rm global\ sky}) < 0.065$.
Of course, all of these galaxies easily pass the magnitude limit,
and the difference between local and global sky subtraction is
smaller for the (far more numerous) smaller galaxies with $r^* > 16$.
We therefore expect negligible systematic bias in target selection
associated with sky subtraction, though there remains a small bias in the
magnitudes of large galaxies much brighter than the survey limit.
For target selection, we adopt the magnitudes of all galaxies measured using
the local sky measurement.

\section{The Galaxy Spectroscopic Target Selection Algorithm}
\label{sec:algorithm}
\begin{figure}
\plotone{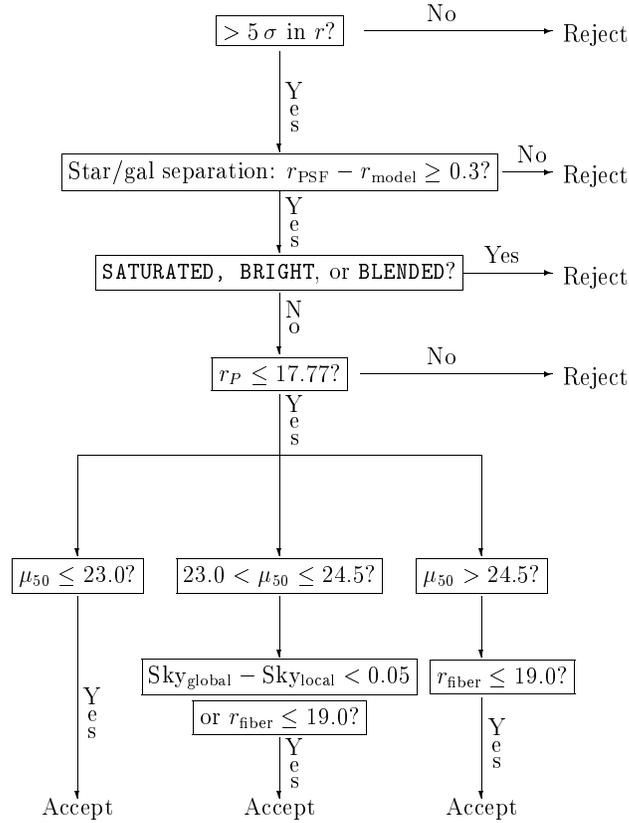}
\caption{Schematic flow diagram of the main galaxy target selection
algorithm.  
All quantities are measured in the $r$ band and are corrected for
foreground extinction. See the text for a full description of all
quantities referred to in this figure. 
}
\label{fig:flowchart}
\end{figure}

\begin{figure}
\plotone{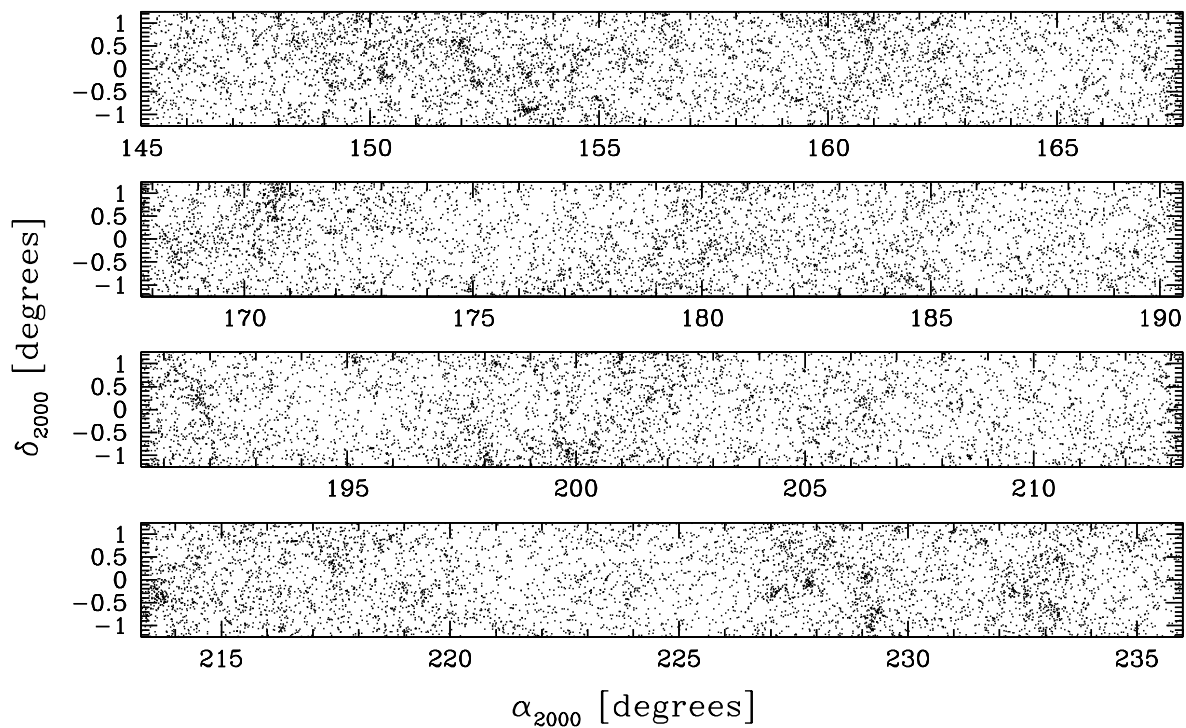}
\caption{The distribution of objects on the sky targeted by the galaxy target
selection algorithm, in a stripe $2.5^\circ$ wide centered on the
Celestial Equator.  We have spectra of essentially all these objects.
The stripe is 91 degrees long, $145^\circ < \alpha
(2000) < 236^\circ$, and is broken into contiguous pieces for the figure.
}
\label{fig:skydist}
\end{figure}

The galaxy target selection 
algorithm is shown schematically as a flowchart in
Figure~\ref{fig:flowchart}. 
The details of this algorithm have been
fine-tuned largely from imaging and spectroscopic observations of a
$2.5^\circ$ wide stripe, $91^\circ$ degrees long, centered on the
Celestial Equator in the Northern Galactic Cap (Runs 752 and 756), 
observed during the
commissioning period of the SDSS \citep{EDR}.  These are the data for
which galaxy counts were presented by \citet{Yasuda}, and the galaxy
luminosity function was presented by \citet{Blanton}.  The
distribution on the sky of galaxies selected by the algorithm
is shown in Figure~\ref{fig:skydist}.  Large-scale
structure is of course apparent.  The SDSS imaging data in this stripe
are taken in a series of twelve parallel {\em scanlines}; note that
the edges between scanlines are {\em not} apparent, which shows
qualitatively that the imaging data are calibrated, and galaxy
selected, consistently.  See \citet{Scranton} for a more detailed and
quantitative discussion of this important point.

\subsection{Magnitude Limit}
\label{sec:maglimit}
We select galaxy targets only from those objects that are detected in 
the $r$-band images, i.e., which are more than 5$\sigma$ above the
sky after smoothing with a Point Spread Function filter.
Before the selection criteria are applied, the photometry is corrected for
Galactic extinction, using the reddening maps of SFD.  
The Petrosian aperture is
unaffected by extinction in the absence of noise (\S~\ref{sec:petrosian}), so the extinction
correction is trivial:
\begin{equation} 
r_P \rightarrow r_P - 2.75 \times E(B-V),
\end{equation}
where the factor of 2.75 converts from the $E(B-V)$
reddenings reported by SFD to the $r$ filter shape, assuming a $z=0$
elliptical galaxy spectral energy distribution.  
Typical extinction values over the SDSS footprint 
lie in the range $ 0 \leq 2.75\times E(B-V) \leq 0.15$. 
There is an effort within the SDSS collaboration to measure
the reddening independently of SFD, using the colors of halo stars,
galaxy counts, and galaxy colors.  This map will be used {\it a
posteriori} to derive a more accurate 
angular selection function, but we anticipate that errors in
the selection function associated with uncertainties in the SFD
map (estimated by SFD to be 15\% of the extinction itself) are already very small.

Our goal is to target a mean of 90 galaxies per square degree in the main
galaxy sample.  This corresponds
to a depth at which the variations of galaxy numbers due to
large-scale structure are quite substantial on degree scales, as
Figure~\ref{fig:skydist} shows.  These fluctuations are consistent
with the measured angular correlation function of galaxies 
(\citealt{Yasuda,Scranton,Connolly}).  
Because of these fluctuations, we need to average over a large area
of sky in order to find the magnitude limit that yields
90 galaxies per square degree.
We have carried this out over 492 square degrees of SDSS
imaging data from a variety of recent SDSS imaging runs.  We have decided on a limiting
magnitude of $r_P = 17.77$, which yields 92 galaxies per
square degree in this region.

Although our selection boundary in apparent magnitude is formally
a sharp one, in practice it is blurred (relative to that of a survey
with no measurement errors) by uncertainties in the reddening
corrections and the inevitable random errors in
magnitude determinations (cf., the discussion in
Appendix~\ref{sec:overlap-integral}).  These errors need to be taken
into account in statistical analyses such as the luminosity function.

\subsection{Star-galaxy Separation}
\label{sec:star-galaxy}
Star-galaxy separation in the SDSS is described by \citet{Scranton}.
In brief, \photo\ generates a detailed model of the PSF at each point in each frame in each band;
this is used as a template to determine a PSF magnitude in each band,
aperture-corrected to an aperture of radius $7.4''$.  In
addition, each object is fit in two dimensions using a sector fitting 
technique (Appendix~\ref{sec:SB}) to a de Vaucouleurs and
an exponential profile of arbitrary axis ratio and orientation, each
convolved with the PSF, and aperture-corrected so that the model and
PSF magnitudes of stars (for which the model scale size will approach
zero) are equal in the mean.  Each of these fits has a goodness of fit
associated with it; the total magnitude associated with the better-fit
of the two models is referred to as the ``model'' magnitude.  
A galaxy target is defined as an object for which
\begin{equation} 
\Delta_{\rm SG} \equiv \rpsf - \rmodel \ge 0.3\ ;
\label{eq:star-gal}
\end{equation}
note that this separation is done at a somewhat more conservative cut
than is done for the star-galaxy separation in \photo\ itself. 

\begin{figure}
\plotone{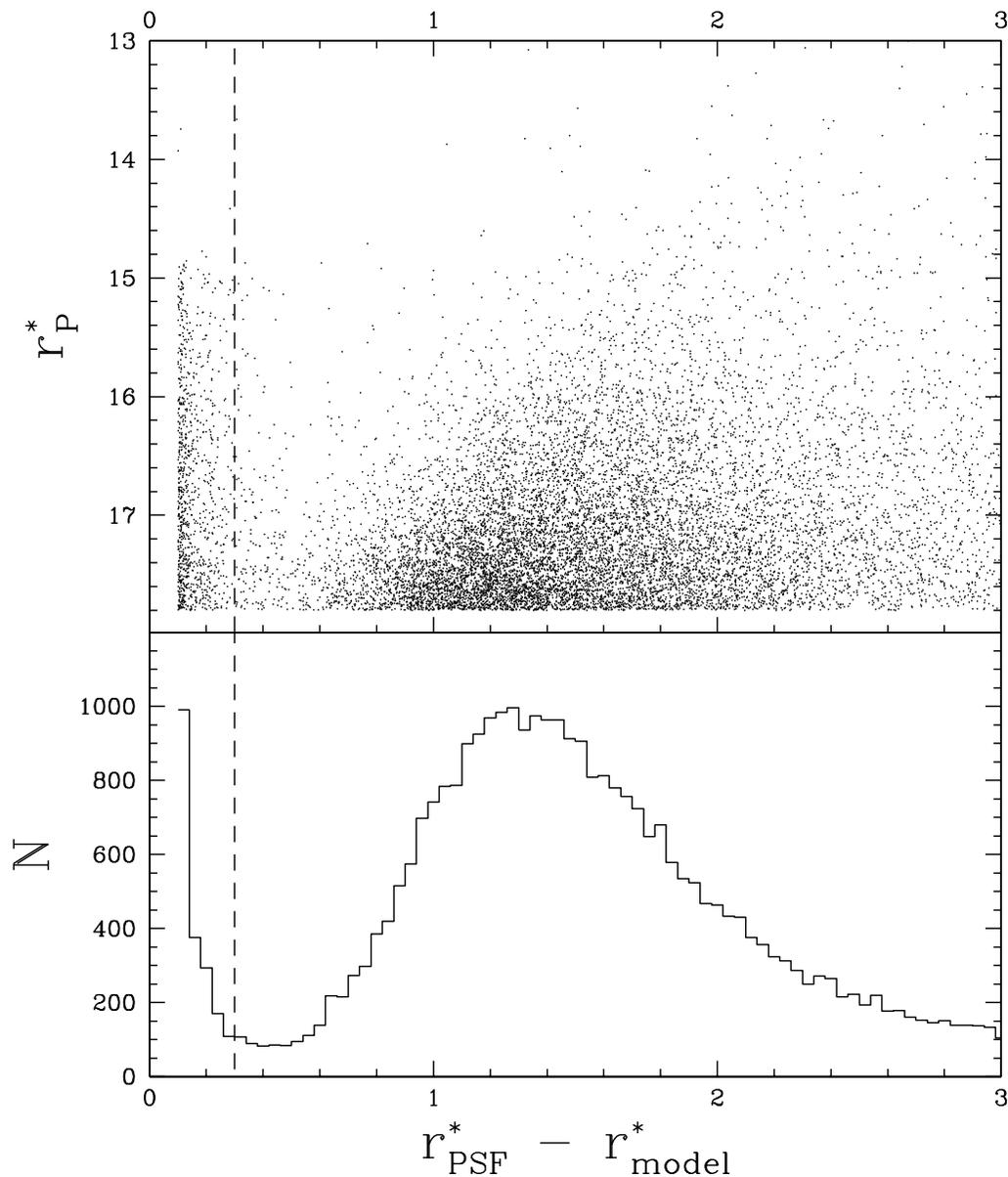}
\caption{The relationship between Petrosian magnitude corrected for
Galactic extinction, and the difference between
PSF and model magnitude in the $r$ band, for stars and galaxies in 115
square degrees.  The separation between stars and galaxies is apparent
with a simple cut in the difference between PSF and model magnitude.  The
distribution of this difference for objects brighter than $r^* = 17.8$ is shown
in the lower panel. 
}
\label{fig:star-gal}
\end{figure}

Figure~\ref{fig:star-gal} shows the distribution of Petrosian
magnitude corrected for extinction as a function of the PSF-model
magnitude difference for 13772 
objects brighter than $r^*_P = 17.8$ over 115 square degrees 
imaged at seeing better than $1.8''$; the marginal distribution of the 
magnitude difference is shown as a histogram on the bottom panel.
At these relatively bright
magnitudes, the distinction between stars and galaxies is very clean,
and there is no evidence for a large population of extremely compact
galaxies that could masquerade as stars.  We discuss the number of stars
that masquerade as galaxies in \S~\ref{sec:stargaltest}. 

\Photo\ models the change of the PSF on scales significantly smaller
than the frame size of $10'\times 13'$.  However, if the seeing
changes rapidly enough, \photo\ cannot estimate a sufficiently
accurate PSF, and the star-galaxy separation suffers.  Such data are
declared not to be survey quality, and are not targeted for spectroscopic
observations
(see the discussion by \citealt{EDR}).  Any resultant holes larger than one
hour in length ($> 15^\circ$) are marked for
reobservation.  Stretches of poor data quality were not uncommon in early SDSS
commissioning data (\citealt{EDR}), but recent improvements to the thermal
environment of the survey telescope have made them increasingly rare.

\subsection{Photometric Flags}
\label{sec:flags}

As described in \S\ref{sec:systematics}, objects detected at more
than $200\,\sigma_{\rm sky}$ have two entries in the database.  To avoid
duplicate targeting of galaxies, target selection therefore rejects
one of these two entries, that flagged as {\tt BRIGHT}.  
The deblending procedures employed by \photo\ are described
briefly by \citet{EDR}; these procedures effectively handle star-galaxy
blends and even galaxy-galaxy blends in the great majority of
cases.  We only target objects that are isolated, 
or children of deblends, or 
parents that are not deblended for one reason or another.
That is, all objects that are flagged as blended, and whose children
are in the catalog, are rejected.  

The vast majority of objects that include saturated pixels are stars,
even if they satisfy the star-galaxy separation criterion of 
equation~(\ref{eq:star-gal}).  We therefore reject
all objects flagged as {\tt SATURATED} in $r$.  We expect no more than
a handful of nearby galaxies, all with bright active nuclei, to have
saturated centers in the SDSS data.  However, our procedure also
rejects a small number of real galaxies that are blended with a
saturated star, since the {\tt SATURATED} flag is passed onto all children of
a parent with saturated pixels if the footprint of the child includes
the saturated pixels.  
We show in \S\ref{sec:completeness} below that the sample
incompleteness introduced by blending of galaxies with saturated
stars is very small, less than 0.5\%.

\subsection{Surface Brightness Limits}
\label{sec:surface-brightness}
\begin{figure}
\plotone{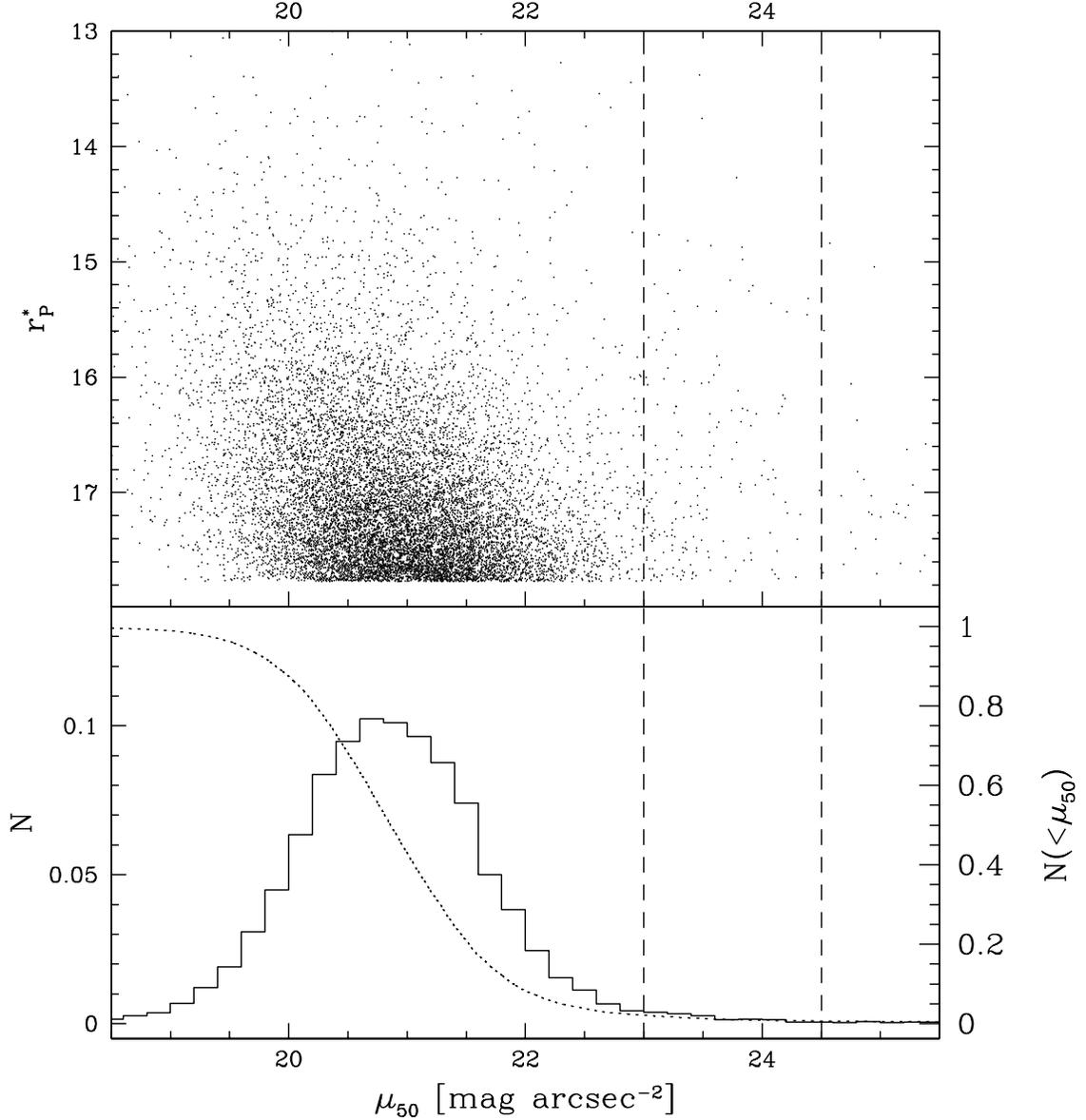}
\caption{The distribution of objects classified as galaxies 
with $r_P^* \leq 17.77$ in the de-reddened, $r$-band, Petrosian
magnitude-Petrosian surface brightness plane.
The cuts at $\muh = 23.0$ and 24.5 are indicated. The lower panel
shows the differential (histogram; left-hand axis) and cumulative
(dotted curve; right-hand axis)
distribution of surface brightness;
of order 1\% of galaxies brighter than the magnitude limit
have $\muh > 23.0$.
}\label{fig:magSB}\end{figure}

As noted by \citet{Shectman} and others, every redshift survey
has at least an implicit surface brightness cut caused by 
detection limits of the imaging data used to derive the input catalog,
and by the limit at which the spectroscopic observations no longer 
yield reliable redshifts.  In order to make this cut simple and
deterministic, we explicitly impose a surface brightness cut ourselves. The
distribution of objects classified as galaxies in the
magnitude-surface brightness plane is shown in Figure~\ref{fig:magSB}.

We target all galaxies brighter than our magnitude cut that have 
half-light surface brightness
$\muh \le 23.0\;\sbunits$ in $r$; visual inspection shows that
essentially all objects down to this surface brightness limit are real
galaxies.  This cut already includes 99\% of all galaxies brighter than our
magnitude limit.  We have visually inspected about 700 lower surface
brightness objects in the range
$23.0 \le \muh \le 26.0$ distributed over 500 square degrees
of imaging data.
In the range $23.0 < \muh < 24.5$, about $65\%$ of these galaxy target
candidates are in fact faint fragments of bright
galaxies (usually spiral arms), cluster cores, or diffraction spikes
of bright stars, erroneously pulled out by the deblending algorithm.
Targeting fragments of bright galaxies does more than waste a
spectroscopic fiber; because two fibers cannot be closer than $55''$,
it can cause the {\em nucleus} of 
the galaxy (which will be a legitimate galaxy target itself) not to be
observed.   
We have found, however, that such spurious galaxy targets have a
local sky biased upwards by the parent
galaxy, and we can therefore
reject them by putting a cut of 0.05 magnitudes on the
difference between the local sky and the mean global sky value 
measured on the frame (cf., the discussion in \S~\ref{sec:systematics}).
Thus, objects in this surface brightness range are
targeted only if the local and global sky values agree to 0.05
magnitudes per square arcsec. 
This algorithm rejects most of the contaminants while rejecting
very few genuine low surface brightness galaxies (LSBs); the fraction
of remaining objects with $23 < \muh < 24.5$ that are
not real LSBs by visual inspection drops to $35\%$.  

  At surface brightnesses fainter than $\muh = 24.5$, we find only 4
real LSBs (out of about 100 candidate objects) distributed over 500 square
degrees, with no obvious automatic way to distinguish them from the
much more common 
ghost images arising due to reflections of bright stars inside the camera.
Moreover, such objects have fiber
magnitudes of order $\rfiber = 21$ or fainter, making it unlikely
that we would be able to measure a successful redshift.  We therefore
do not target these objects.   
Low surface brightness objects tend to be of low intrinsic luminosity, 
so they are visible only within a small volume, and their rarity in
the sample does not necessarily translate into a small volume density;
however, they probably contribute
little to the overall luminosity density of the universe
(see the discussions by \citealt{Blanton,Cross}).
Since we find only four galaxies with $\muh > 24.5$ in an area that
contains 45,000 main sample galaxies, we estimate that only $\sim 0.01\%$
of galaxies brighter than our magnitude limit are rejected by our
surface brightness cut.

For very nearby galaxies ($cz < 10,000 \rm\,km\,s^{-1}$), the Petrosian half-light radius can be
quite large, substantially larger than the $3''$ aperture of the
fibers.  If a low-surface brightness galaxy is strongly nucleated,
our surface brightness cut can result in missing a potentially
interesting galaxy that would easily yield a redshift.
We therefore accept objects of any surface brightness brighter than
our Petrosian magnitude limit, 
if their fiber magnitude in $r$ is brighter than 19.0.  In practice,
very few objects enter the sample this way; for example,
the stripe shown in Figure~\ref{fig:skydist} contains
no galaxies targeted in this manner.

There is non-negligible cross-talk between adjacent fibers in the
spectrographs, thus overly bright objects make the extraction of
the spectra of neighboring faint objects difficult.  To avoid this, we
reject objects whose fiber magnitude 
is brighter than 15 in $g$ and $r$, and 14.5
in $i$.  This criterion rejects about $0.07\%$ of real galaxies
that would otherwise be included in the galaxy spectroscopic sample.

Finally, we reject all objects brighter than $r_P=15.0$ that have $\Rh <
2''$.  This cut rejected a small number of bright stars that managed to 
satisfy equation~(\ref{eq:star-gal}) during the commissioning phase of the 
survey, when the star-galaxy separation threshold was  $\Delta_{SG} = 0.15$.
With the final value of the cut at $\Delta_{SG} = 0.3$,
there are no objects in Runs 752 and 756 that are rejected by this
cut alone. However, we still enforce this criterion, since such small, bright
objects will saturate the spectroscopic CCDs and contaminate
the spectra of adjacent fibers, and they are in any case
more likely to be star-galaxy separation errors than actual galaxies.

\subsection{Tiling and Fiber Assignment}
\label{sec:tiling}

Once targets have been selected, they are {\em tiled}, i.e., 
assigned to spectroscopic plates in a way that optimizes the observing
efficiency \citep{tiling}.  This
process leaves essentially no systematic spatial gaps in the
distribution of tiled objects, except at
the outer boundary of the tiled region (where they will be tiled in
a subsequent run based on more imaging data).

\begin{figure}
\plotone{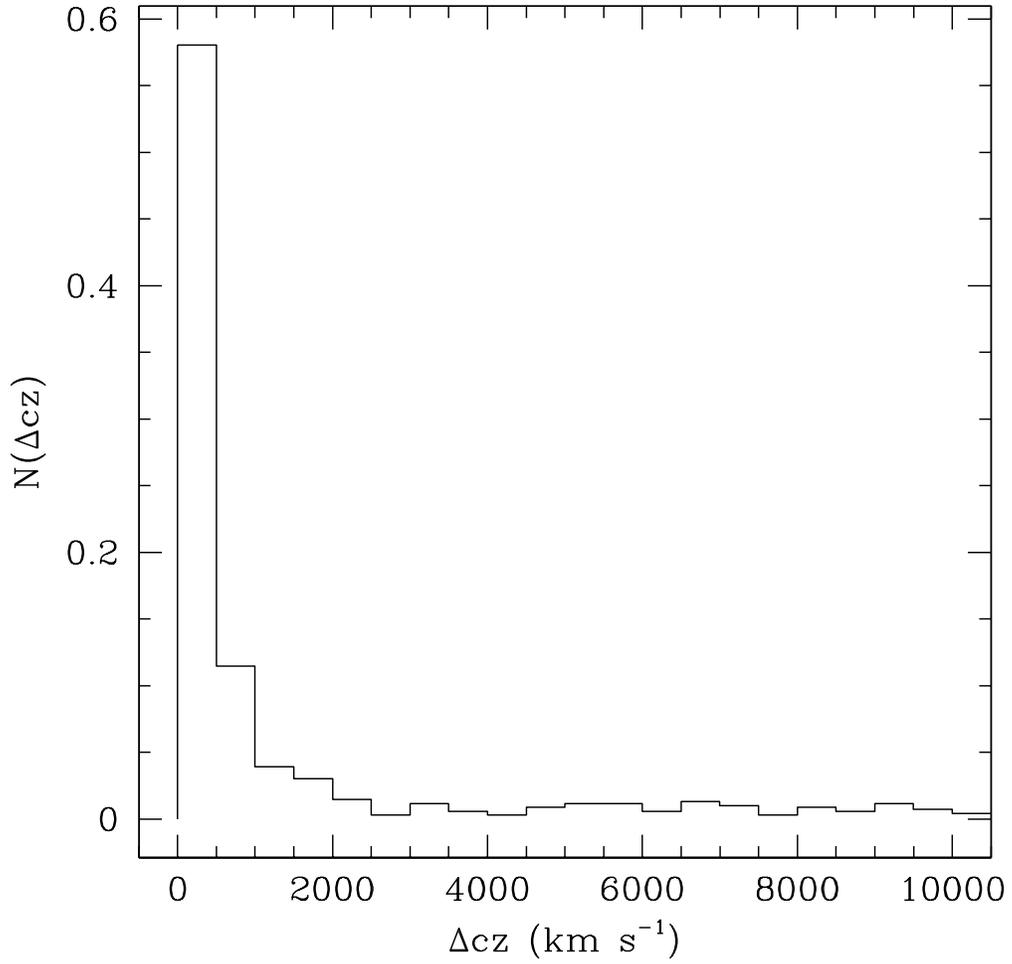}
\caption{The distribution of redshift differences of pairs of galaxies
separated on the sky by less than $55''$, as measured from regions in
which two or more spectroscopic plates overlap.  58\% of pairs have a
redshift difference less than 500 km s$^{-1}$.}
\label{fig:fiber-collision}
\end{figure}

However, as mentioned earlier, fibers cannot be placed closer than
$55''$, center to center, on a given plate.   The tiling algorithm
maximizes the number of targeted objects given this restriction. 
The tiled objects in decreasing order of priority are
brown dwarfs and hot standards (both very rare, of order one object
per plate), quasar candidates (of order 100 per plate), and finally 
(at equal priority, and composing the bulk of the objects) the LRGs
and main sample galaxies.  
An object with a higher priority will never lose a fiber in favor of
a lower priority object.
However, for two objects of equal priority --- in particular, for
two main sample galaxies --- the choice of observed target is
made at random.  In regions where plates overlap (roughly 30\% of the
tiled region), both members
of a close pair are often observed.  Figure~\ref{fig:fiber-collision}
shows the measured redshift difference between such pairs
(\citealt{Zehavi}); 58\% of 
close pairs have a redshift difference less than 500 km s$^{-1}$ (compared
with only 4\% of pairs of arbitrary angular separation).  
We note, however, that pairs in overlap regions may not be representative
of the full pair population, since the locations of overlap regions
are influenced by the galaxy clustering pattern.

\citet{Zehavi} show that for large-scale structure statistics, it is
often sufficient to simply double-weight the observed galaxy of a
close pair in statistical calculations, or to assign the unobserved
galaxy the same redshift as the observed one (which is similar to
double weighting but uses the angular position of the unobserved
object).  
Roughly 6\% of
all galaxy targets are not assigned a fiber due to the $55''$ restriction.

\section{Tests of Algorithm Performance}
\label{sec:test}

In this section, we present tests of the target selection algorithm.
We first show (\S~\ref{sec:stargaltest}) that our star-galaxy
separation works well; less than 2\% of the galaxy targets turn out to
be stars, while less than 0.5\% of true galaxies are rejected by our
algorithm.  This leads into a discussion of the
spectroscopic characteristics of the sample, \S~\ref{sec:spectra}.  
We then carry out
various tests of the completeness of the selection
(\S~\ref{sec:completeness}), and find that the sample completeness exceeds 
99\%, though it becomes somewhat lower for brighter galaxies, which
are more likely to be rejected because of blending with saturated
stars.  Finally,
\S~\ref{sec:reproducibility} uses repeat scans of an extended area of
sky to quantify the reproducibility of the algorithm; the differences
in targeted objects are consistent with expectations due to
random photometric errors.

The SDSS science requirements for the galaxy sample include
completeness of at least 95\%, a redshift success rate of at least
95\%, stellar contamination of less than 5\%, and insensitivity of
selection to observing conditions during imaging.  The tests below
demonstrate that the galaxy sample easily satisfies these
requirements.

\subsection{Tests of Star-Galaxy Separation}
\label{sec:stargaltest}

We select as galaxy targets objects with $\Delta_{SG} \equiv r_{PSF}
- r_{model} \ge 0.3$ (\S~\ref{sec:star-galaxy}).  
During the commissioning phase of the survey, we selected galaxy targets
using a more permissive value of the star-galaxy separation
threshold at $\Delta_{SG} = 0.15$. 
About 3\% of roughly 6000 galaxy targets selected from
imaging data with seeing better than $1.8''$ lie in the range 
$0.15 < \Delta_{SG} < 0.6$.  Of these, only 10\% of the targets with
$\Delta_{SG}< 0.3$ are actually galaxies from their spectra, while
the galaxy fraction rises to 20\% for those with  $0.3 < \Delta_{SG}<
0.45$ and to 65\% for targets with  $0.45 < \Delta_{SG}< 0.6$. 
The major contaminants at lower values of $\Delta_{SG}$ are single stars, while
the contaminants above $\Delta_{SG} = 0.3$ are mostly double stars too
close ($<3''$) to be deblended into single stars by \photo.
The stellar contamination to the galaxy sample is quite independent
of seeing as long as the seeing full width at half maximum is smaller
than $1.8''$ and does 
not change rapidly (i.e., data of survey quality; see
\S~\ref{sec:star-galaxy}).  

With our final cut of $\Delta_{SG} > 0.3$, slightly under 2\% of
galaxy targets are single and double stars. A similar fraction of the targets are
classified spectroscopically as quasars; these of course are successes
of the algorithm, as they are mostly low-redshift AGN. 
The spectroscopic results from the $\Delta_{SG}=0.15$ threshold test
run imply that 
only about 0.3\% of true galaxies brighter than our magnitude limit
are rejected by our star-galaxy separation criterion.

  Finally, the SDSS quasar target selection algorithm \citep{QSO}
targets stellar objects with colors distinct from those of ordinary stars.  As
the colors of galaxies are distinct from those of individual stars
\citep{Strateva}, unresolved galaxies are likely to be selected by
that algorithm.  As \citet{QSO} demonstrate, some star-forming
galaxies and E+A galaxies are indeed selected, but only a handful of
such objects brighter than $r^*_P = 17.77$ are not already selected by
the galaxy target selection algorithm; these objects represent less
than 0.1\% of the galaxy sample.  Of course, unresolved quasars, which
are not included in this total, can be brighter than $r_P^* = 17.77$;
such objects are explicitly selected by the quasar target selection
algorithm.

\subsection{Spectroscopic Characteristics}
\label{sec:spectra}

\begin{figure}
\plotone{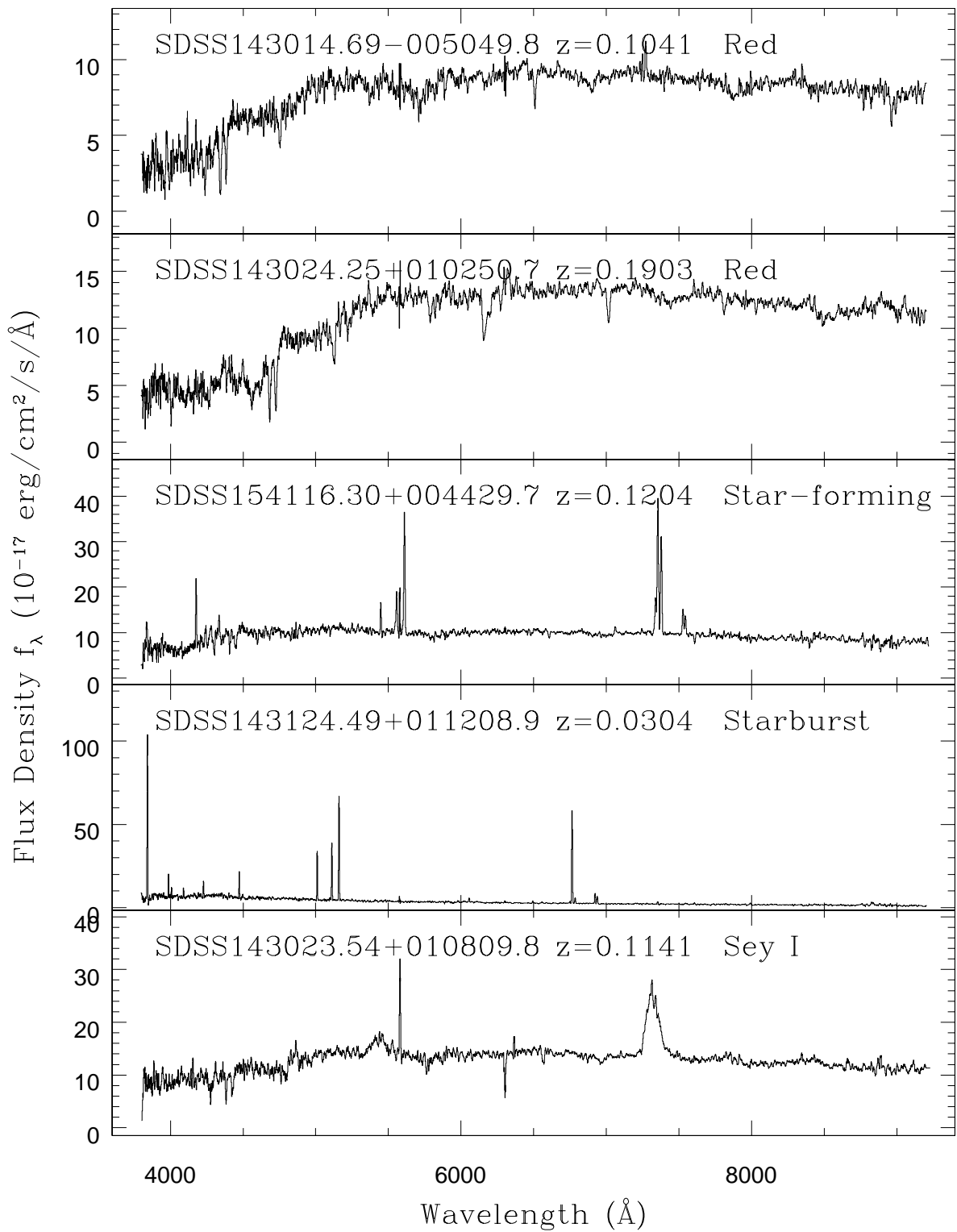}
\caption{
Representative SDSS spectra of galaxies from the sample, smoothed with
a 5-pixel boxcar.  These
objects are all close to the magnitude limit of $r^*_{Petro} =
17.77$.  Shown are two red galaxies, a star-forming galaxy, a
starburst galaxy, and a Seyfert 1 galaxy.}
\label{fig:spectra}
\end{figure}

Figure~\ref{fig:spectra} shows
representative SDSS spectra of galaxies close to our photometric limit.
The S/N values of the spectra are determined by their fiber
magnitudes; the $r$ fiber magnitude distribution for the targetted sample
is centered at $r_{fiber} = 18.5$; essentially all galaxies have fiber
magnitudes brighter than 20 (at which the spectroscopic S/N per pixel
is required to be greater than 4; \S~\ref{sec:SDSS}).  Thus the fraction of galaxy targets whose spectral
S/N is not adequate to measure a redshift is less than 0.1\% of
the total main galaxy targets.  The objects in the tail of the fiber
magnitude distribution (at $r_{\rm fiber} > 20.5$) are low surface
brightness galaxies.  The SDSS spectra are 
consistently of good enough quality to yield reliable redshift
measurements (i.e., with statistical error $< 30\rm \, km\,s^{-1}$) for
galaxies at fiber magnitudes as faint as $r_{\rm fiber} = 20.8$.
Thus the impact of spectroscopic observing conditions
on sample completeness is essentially negligible.

\begin{figure}
\plotone{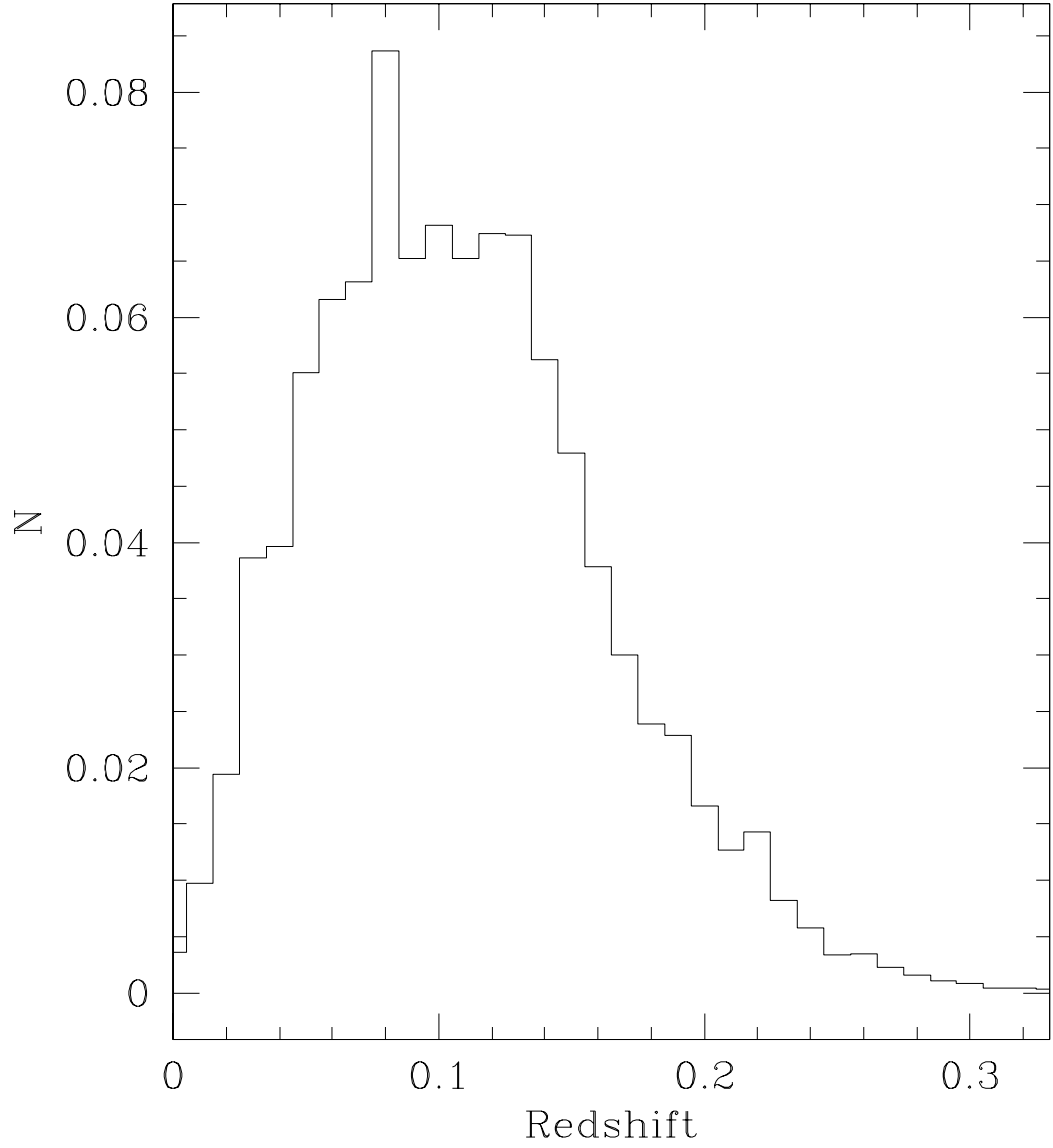}
\caption{The redshift histogram of 57,366 galaxies selected by
the algorithm; the y-axis is the number of galaxies in bins of 0.01 in
redshift.  Even this large number of galaxies is not quite enough to
average over large-scale structure fluctuations.} 
\label{fig:zhist}
\end{figure}

Figure~\ref{fig:zhist} shows the redshift histogram for 57,366
galaxies observed in 2000 and 2001.  The 
median redshift is 0.104; even with this large number of galaxies, the
redshift distribution is not completely smooth, due to large-scale
structure fluctuations. A pie diagram of the galaxy distribution in
redshift space from the sample is shown by \citet{Zehavi}. 

\subsection{Completeness of the galaxy sample}
\label{sec:completeness}

We anticipate
the most serious problems with completeness of the galaxy sample at the bright end, where
galaxy photometry can be the most problematic; big galaxies often have substructure (bars, spiral arms, dust
lanes, HII regions, and so on), which can fool an automated deblender.  In this subsection, we
compare the SDSS galaxy sample with samples drawn from the Two Micron
All-Sky Survey (2MASS)\footnote{\url 
\tt http://www.ipac.caltech.edu/2mass/overview/about2mass.html}, the Zwicky catalog, and visual inspection of
SDSS images, and show that our completeness is of order 99\% overall,
worsening to 95\% for galaxies brighter than $r^*=15$. 
Here we define completeness to be the fraction of galaxies satisfying
our selection criteria that are in fact identified as spectroscopic
targets by the automated algorithm.  About 6\% of these targets are
not observed spectroscopically 
because of the $55''$ fiber separation constraint (see \S\ref{sec:tiling}).
However, the locations of these missed galaxies are known,
and they can be accounted for in any statistical analysis
of the galaxy population, so we do not count them as contributing
to incompleteness.  

Falco \etal\ (1999) have compiled accurate astrometry for galaxies in the
Zwicky \etal\ (1961-68) catalog; we have matched this list with the SDSS
database.  This catalog is limited to $b_{\rm Zwicky} = 15.7$,
corresponding roughly to $r^* \approx 15$.
Roughly 90\% of the 176 Zwicky galaxies in 77 square degrees of sky
have a corresponding SDSS main galaxy target within $3''$ of the
nominal Zwicky position.  
Visual inspection shows that the galaxy astrometry in the Zwicky catalog
is inaccurate enough to cause a mismatch for about $5\%$ of the
galaxies; all of these galaxies in fact have an
SDSS spectroscopic galaxy target centered at the correct
photometric centroid of the galaxy.
The remaining 5\% of the galaxies in the Zwicky catalog have corresponding
SDSS galaxy candidates that are flagged as {\tt SATURATED} due to
overlap with a saturated star (\S~\ref{sec:flags}), and hence these
galaxies will not be targeted spectroscopically. 
Thus
of order 95\% of galaxies in the Zwicky catalog are
being targeted spectroscopically.  As one goes fainter, the fraction of
galaxy targets that are missed due to the {\tt SATURATED} flag will go down,
since the area covered by the galaxies becomes smaller.

Yasuda \etal\ (2001) visually inspected all objects brighter than $r^*
= 16$ in the stripe defined by Runs 752 and 756 (\citealt{EDR}), and
classified them into stars and galaxies. There are 1743 galaxies in
this sample over 200 square degrees of sky, roughly 10\% of the full
galaxy sample to $r^*=17.77$ in this region.
The galaxy target selection algorithm selects 1701
$(97.6\%)$ of these bright galaxies for spectroscopic observations.
Of the 42 galaxies that are not targeted, 30 are blended with saturated
stars and hence have the {\tt SATURATED} flag set, while the remaining 
12 galaxies are rejected because they 
have fiber magnitudes brighter than $i^* = 14.5$.

Finlator \etal\ (2000) discuss the matching of the SDSS
photometric catalog with that of 2MASS.  
Given typical  colors of galaxies of $r-K = 3.0$,
the SDSS spectroscopic magnitude limit corresponds to  $K \approx
14.7$, which is comparable to the photometric limit of 2MASS.  
Of order 2\% of 2MASS point sources in the region
of overlap of the two surveys are not found in the SDSS database; of
these, 2/3 are asteroids, and the remaining 1/3 do not enter the
SDSS catalog because they are associated with complicated blends
often involving saturated stars or diffraction spikes which are not
deblended properly 
\citep{Ivezic01}.  Assuming that the latter 1/3 of the missing 2MASS sources 
are real, the lower limit of the completeness
of the SDSS galaxy catalog at the spectroscopic limit is 99.3\%. 

Finally, 
we performed an end-to-end study of 
the completeness of the spectroscopic galaxy sample by visually inspecting 
all objects brighter than $r^* = 18$ 
and with $\rpsf - \rmodel \ge 0.1$ (and with no other cuts based 
on flags) over 22 square degrees of sky.
These two limits are more permissive
than the corresponding cuts in the galaxy target selection algorithm,
thus allowing us to quantify the number of real galaxy targets that
we miss by these sharp cuts. 
Note that this test is complete to the extent that \photo\
successfully finds all objects brighter than $r^*=18$ in the first
place.  This is more than four magnitudes brighter than the SDSS
``plate limit'', and indeed, tests of the repeatability of found objects in regions of sky
observed by SDSS more than once show that essentially 100\% of objects
at this brightness are found both times. 

There are 3186 unique entries in our visual inspection sample.  Of
these, 
366 are either single or double stars by visual inspection (the
vast majority of which do not satisfy our star-galaxy separation
criterion, of course).
In addition, 464 objects have dereddened Petrosian magnitudes fainter than
$r^* = 17.77$. 
This leaves 2356 unique objects
that should have been targeted by the selection algorithm.  Of
these, we target 2330 (98.9\%).  We have spectra of 2184 of these; 21
are classified as quasars spectroscopically, 26 as stars (and thus are errors
in our visual classification), two were
satellite trails in the imaging data, and the
remainder are galaxies.  Nearly all the 146 objects for which we have not
obtained spectra (6\%) have companions within $55''$.  It
occasionally happens that a spectroscopic fiber will break or fall
out, resulting in the absence of a spectrum despite a hole being
drilled for it; there was only a single example of this problem among the
2330 spectroscopic targets in this sample. 

 The $2356-2330=26$ objects that we do not target break down as follows.  We miss
10 galaxies because they are blended with saturated stars, while another 6 galaxies do not pass our
star-galaxy separation criterion. We miss another 10
galaxies due to an error in setting the flags in \photo\ (version 5.2
and earlier). This error
has been corrected in V5.3 of \photo, and we expect not to
lose these galaxy targets in the future. 
Thus, this end-to-end test implies that only $10/2356=0.4\%$ of galaxies
in this 22 square degree area that should be targeted would not be,
with the corrected version of \photo, and that $6/2356=0.25\%$ of
true galaxies would be rejected by our star-galaxy separation criterion.
The intrinsic sample completeness therefore exceeds 99\%.  The only significant
cause of incompleteness that we have identified is blending with 
saturated stars, affecting only 5\% of bright galaxies, and
a negligible fraction fainter than $r_P = 16$. 

Ten percent of sample galaxies have a neighbor within $55''$.  The
fraction of these missed galaxies that will be recovered in a
subsequent observation of an overlapping plate depends to some extent
on the size and geometry of regions that are tiled for spectroscopy,
but the estimate from this 22 square degree region that $\sim 6\%$ of
galaxy targets will ultimately remain unobserved appears reasonable
and consistent with our more recent experience.

\subsection{Reproducibility}
\label{sec:reproducibility}

The SDSS scanlines overlap by about $1'$ at their edges; moreover,
we have several scans that cover the same area of sky.  This allows us
to test whether we get consistent results of photometry
and main galaxy target selection in the overlaps.

We have tested the repeatability of the galaxy target selection
algorithm by selecting galaxy targets from repeated scans of the
same region of the sky. In particular, the SDSS imaging runs 745 
(observed on Mar 19, 1999) and 756 (observed on Mar 21)
scanned the same patch of sky 
($160.5 < \alpha < 235.5$ on the Celestial equator); the
six columns of the two runs spanned the same range in declination.

Figure~\ref{fig:magerr} shows the difference in the 
$r$-band Petrosian magnitudes of galaxies in common between these runs 
brighter than $r^*=18$. 
There is no offset in the mean of the two measurements of the 
Petrosian magnitudes of these galaxies, and the rms differences in the
$r$-band is 0.035 mag, in good agreement with the estimated Petrosian
magnitude errors at the sample magnitude limit.  

Over a $90$ deg$^{2}$ region of repeated imaging data, we select
9159 (9125) galaxy targets from Run 745 (756). Of these, 8652 (94.5\%)
targets have a corresponding target within $0.7''$ in the other run.
There are another $57$ (0.6\%) targets in Run 745 that have a 
corresponding galaxy target within $3''$ in Run 756.
Of the remaining 450 targets in Run 745 that are not selected
in Run 756, 342 objects (3.7\%) have a corresponding object within
$0.7''$ in Run 756 that is fainter than the magnitude limit.
This fraction is comparable to the fraction of galaxies
that is expected to cross the magnitude limit in two repeated scans
because of random
photometric errors, as discussed in Appendix~\ref{sec:overlap-integral}.
Another 78 $(0.9\%)$ targets in Run 745 are rejected by  the star-galaxy 
separation algorithm in Run 756, while 30 objects $(0.3\%)$ are 
saturated in the Run 756 images.  The discrepancy in
star-galaxy separation is significantly worse than we quoted in
\S~\ref{sec:stargaltest}, due to the fact that Run 745 had seeing
significantly worse than our criterion for survey-quality data.
The fraction of targets selected from 
Run 756 but not from Run 745 follows similar statistics.    Thus
the repeatability of the galaxy target selection sample is probably
better than 95\%, and nearly all of the non-repeatability can be
attributed to expected random photometric errors, which should not introduce 
any systematic biases in statistical studies.

\section{Conclusions}
\label{sec:conclusions}

\subsection{Summary of the algorithm performance}

The main spectroscopic galaxy sample of the SDSS is a
reddening-correct $r$-band magnitude 
limited sample of galaxies brighter than $r_{P} = 17.77$, with an estimated 
surface density of 92 galaxies per square degree.
The magnitude is measured within a Petrosian aperture, so as to provide
a meaningful measure of a fraction of the total light of the galaxy that
is independent of distance to the galaxy, reddening, and sky background.
Star-galaxy separation is based on the difference between PSF and
galaxy model magnitudes, which effectively quantifies the extension of
the source relative to a PSF.  We reject objects with Petrosian half-light
surface brightness $\muh > 24.5$, a cut that eliminates $\sim 0.1\%$
of galaxies brighter than the magnitude limit.
In the range $23 < \muh < 24.5$, we use a measure of the difference
between local and global sky brightness to increase our efficiency
of targeting real galaxies.

We have objectively tested the star-galaxy separation algorithm and the
completeness and reproducibility of the spectroscopic sample using
imaging and spectroscopic data taken during the commissioning phase of 
the survey.  During commissioning, we refined the criteria in the target selection algorithm
to achieve our goals on the completeness and efficiency of the 
spectroscopic sample. At the time of this writing, we find that the
star-galaxy separation is accurate to better than $2\%$, with the
main contaminants being close double stars.
The fraction of true galaxies rejected by the star-galaxy
separation criterion is only $\sim 0.3\%$.

The completeness of the main galaxy sample is a function of magnitude.
At bright magnitudes ($r^* < 15$), we find that we target
$95\%$ of the galaxies in the Zwicky catalog, while the remaining $5\%$
are missed because they are blended with saturated stars. From comparison
with visual inspection of bright galaxies ($r^* < 16$)
over 200 square degrees of sky,
we find that the  completeness increases to about $97.6\%$.
Finally, from comparison with a visual inspection of all objects brighter 
than $r^* = 18$ over 22 square degrees of sky, we find that 
the completeness of the
galaxy sample to the magnitude limit is above $99\%$.
The only significant source of incompleteness that we have identified is 
blending with saturated stars; this incompleteness is higher for
brighter galaxies because they subtend more sky.  

Essentially all
main sample galaxies (99.9\%) that are observed spectroscopically
yield successful redshifts.  About 10\% 
of galaxy targets do not receive a fiber on the first spectroscopic pass 
because they lie within $55''$ of another sample galaxy.
Some of these galaxies lie in regions of plate overlap and are
observed subsequently, and the fraction of galaxies that are 
missed in the end because of the fiber separation constraint is about 6\%.  
These missing galaxies can be
accounted for in any statistical analysis by appropriate weighting
of the galaxies in close pairs that are observed.  

We have tested the reproducibility of the galaxy sample by selecting targets
from repeated scans of the same region of the sky. We find that 
$94.5\%$ of the spectroscopic sample galaxies are selected in both the
scans. About $3.7\%$ of galaxies fall out of the sample because
they cross the magnitude limit and are replaced by a similar number
of galaxies crossing in the other direction; this fraction is consistent
with expectations based on random errors in the Petrosian magnitudes.
Other galaxies fall out of the sample because of changes in saturation
or star-galaxy separation.  Reproducibility of target selection is
therefore high, and the random photometric errors that lead to 
non-reproducibility are not expected to cause systematic biases
in statistical analyses.

\subsection{Scientific applications of the SDSS imaging and spectroscopic data}

The imaging data on which we tested and refined the galaxy target
selection algorithm, and the resulting galaxy spectroscopic sample
have been studied in the context of both large scale structure and
properties of galaxies. Extensive tests by Scranton \etal\ (2001) show
that the imaging data obtained by the SDSS are free from internal and
external systematic effects that influence angular clustering
for galaxies brighter than $r^* = 22$,
almost four magnitudes below the limit of the
spectroscopic sample.  At the bright end, \citet{Yasuda} studied the
bright galaxy 
sample in the same data, and showed that the photometric pipeline
correctly identifies and deblends blended objects and provides correct
photometry for bright $(r^* < 16)$ galaxies.

The spectroscopic galaxy sample targeted using development versions of
the target selection algorithm during the commissioning phase of the
survey has been used to measure the luminosity function of galaxies as
a function of surface brightness, color, and morphology (Blanton \etal\
2001). A primary goal of the SDSS is to measure the properties of
large scale structure as traced by different types of galaxies.
Zehavi \etal\ (2002) used this spectroscopic
sample to measure the correlation function and pairwise velocity
dispersion of samples defined by luminosity, color, and morphology.
Bernardi \etal\ (2002) used the spectra and photometry to study the correlations
of elliptical galaxy observables including the luminosity,
effective radius, surface brightness, color, and velocity
dispersion. All these studies show that the galaxies targeted
spectroscopically by the SDSS constitute a uniformly selected sample
spanning a wide range of galaxy types, ideal for 
analyses of large scale structure and galaxy properties.

\acknowledgements Funding for the creation and distribution of the
SDSS Archive has been provided by the Alfred P. Sloan Foundation, the
Participating Institutions, the National Aeronautics and Space
Administration, the National Science Foundation, the U.S. Department
of Energy, the Japanese Monbukagakusho, and the Max Planck
Society. The SDSS Web site is http://www.sdss.org/.

                         The SDSS is managed by the Astrophysical
Research Consortium (ARC) for the Participating Institutions. The
Participating Institutions are The University of Chicago, Fermilab,
the Institute for Advanced Study, the Japan Participation Group, The
Johns Hopkins University, Los Alamos National Laboratory, the
Max-Planck-Institute for Astronomy (MPIA), the Max-Planck-Institute
for Astrophysics (MPA), New Mexico State University, Princeton
University, the United States Naval Observatory, and the University of
Washington.

MAS acknowledges the support of NSF grants AST-9616901 and AST-0071091. 
DHW acknowledges the support of NSF grant AST-0098584 and the
Ambrose Monell Foundation.

\appendix
\newcommand{\XXX}[1]{\textbf{XXX} #1}

\section{The Calculation of Petrosian Quantities by \Photo}
\label{sec:petro-appendix}

\subsection{Measuring Surface Brightnesses}
\label{sec:SB}

The photometric pipeline, \photo, measures the radial profile of every object by measuring the
flux in a set of annuli, spaced approximately exponentially
(successive radii are larger by approximately $1.25/0.8$); the outer
radii and areas are given in Table 7 of \citet{EDR}.  Each annulus is
divided into twelve 30$^\circ$ {\it sectors}.  For the inner six annuli (to
a radius of about 4.6 arcsec) the flux in each sector is calculated by
exact integration over the pixel-convolved image;
for larger radii the sectors are defined by a list of the pixels that
fall within their limits.  Usually the straight mean of the pixel
values is used, but for sectors with more than 2048 pixels a mild clip
is applied (only data from the first percentile to the point
2.3$\sigma$ above the median are used).

Given a set of sectors, \photo{} can measure the radial profile.  If
the mean fluxes within each of the sectors in an annulus are $M_j (j
=1,\cdots,12)$, it calculates a point on the profile (`{\tt profMean}') as
$$
I_i = \frac{1}{12} \sum_{j=1}^{j=12} M_j ~.
$$
The error of this quantity (`{\tt profErr}') is a little trickier.  
If we knew that
the object had circular symmetry, we would estimate it as the variance
of the $M_j$ divided by $\sqrt{12}$.  Unfortunately, in general
the variation among the $M_j$ is due both to noise and to
the radial profile and flattening of the object.  To mitigate this
problem, we estimate the variance as
\begin{eqnarray}
\hbox{Var}_I = \frac{4}{9} \times
	\frac{1}{12} \sum_{j=1}^{j=12} \left[M_j - {1 \over 2}
	  (M_{j-1} + M_{j+1})\right]^2 ~,
\label{EqnProfErr}
\end{eqnarray}
where we interpret `$j \pm 1$' modulo 12, and where the
factor $4/9$ is strictly correct in the limit that all the $\langle{M_j}\rangle$ are
equal. 
This use of a local mean takes out linear trends in the profile around
the annulus, and results in an estimate of the uncertainty in the profile
that is a little conservative, but which includes all effects.  The
error due to photon noise alone, if one needed this, 
could easily be calculated from the $I_i$ and the known gain of the CCD.

In practice, \photo{} doesn't extract the profile beyond the point
that the surface brightness within an annulus falls to (or below)
zero.

\subsection{Measuring the Petrosian Ratio}

The $I_i$ measured in the previous subsection represent the surface
brightness at some point in the $i^{\hbox{th}}$ annulus, but exactly
what point is not clear.  Instead of making some assumption about the
form of the radial profile, we preferred to work with the cumulative
profile, as the $I_i$ (and the known areas of the annuli, $A_i$) define
unambiguous points $C_i$ on the object's curve of growth. By using
some smooth interpolation between these points (which, of course,
makes an assumption about the form of the radial profile) we can
estimate the surface brightness at any desired radius.

The cumulative profile has a very large dynamic range, so in practice
we make this interpolation using a cubic spline on the $\asinh \R$ vs.\
$\asinh C$ curve\footnote{We chose $\asinh$ rather than a logarithm as it is 
well behaved near the origin; cf. \citet{asinhMags}.}, where $\theta$ is the
angular distance from the center of the object.  As with any
cubic spline, we need to specify two additional constraints to
fully determine the curve; we chose to use the `not-a-knot' condition,
i.e. we force the \textit{third} derivative to be continuous at the
second and penultimate points.  We also used a `taut' spline, which
adds extra knots wherever they are needed to avoid the extraneous
inflection points characteristic of splines put through sets of
points with sharp changes of gradient (e.g., two straight line
segments; \citealt{deBoor}).  We explored the use of smoothing splines,
but found that they didn't conserve flux --- it is important
that the curve actually pass through the measured points!

The choice of boundary conditions at the origin is a little tricky. The
gradient \\
$d\, \asinh C/d\, \asinh \R$ is 0 at the origin --- but only
\textit{very} close to the origin. For a constant surface brightness source,
once $C \gg 1$, the gradient becomes very large, scaling as $\approx
1/\R$ for $\R \ll 1$. 
Experiment showed that the best results were achieved by not imposing
any symmetry (and thus gradient) constraints at $\R = 0$.

With the cumulative profile, expressed as a spline, in hand, the
Petrosian ratio is easily calculated, as defined in equation~(\ref{eq:RPetro}).
In practice, we evaluate $\Pratio$ at the annular boundaries
$\R_i$ (where we know the cumulative profile), and use another taut
not-a-knot spline to interpolate $\Pratio$ as a function of $\asinh \R$.

We estimate the uncertainty $\sigma_\Pratio$ in $\Pratio$ by
propagating errors in equation~(\ref{eq:RPetro}) on the assumption of
Poisson noise in the object and sky; we allow for the covariance
between the numerator and denominator and only work to quadratic order
in the errors.  However, if the resulting
S/N exceeds that of the measured
radial profile at that point (measured as described in
eq.~\ref{EqnProfErr}), the error in $\Pratio$ is set to $\Pratio$
times the error in the radial profile. 

\subsection{Measuring the Petrosian Flux}

The Petrosian radius $\RP$ is found by solving the equation
$\Pratio(\RP) = \RPcrit$.  We've expressed
$\Pratio$ as a cubic spline, so we can piecewise apply the usual analytic
formula for the roots of a cubic and
find all Petrosian radii (there may indeed be more than one solution;
see below).  Clearly $\Pratio(0) = 1$, and
for most forms of an object's radial profile $\Pratio(\infty) = 0$
(the exception being a power law $P(\R) \propto \R^{-\alpha}$, with
$\Pratio(\infty) = (\alpha - 2)/\alpha$), so almost all objects will
have at least one $\RP$.

Even in the absence of noise, some objects may have more
than one $\RP$; the Petrosian ratio need not be monotonic in $\R$. For
example, a galaxy with an AGN can have one Petrosian radius for the 
part of the galaxy where its light is dominated by the nucleus and
another, larger, $\RP$ associated with the extended light; in this
case we should adopt the \textit{larger} value.  On the other hand, a bright
star with a much fainter galaxy nearby that has not been properly
deblended can have a small $\RP$
associated with the star, and another much larger value produced
by a small rise in the radial profile at the position of the galaxy,
at a radius where the \textit{mean} enclosed surface brightness due
to the star has fallen to a low value; in this case we should adopt
the \textit{smaller} value.

These spurious values of $\RP$ are found at a point where the
surface brightness is very low, so we have adopted the following
procedure, setting flags for each object to describe any unusual
problems we come across, as described by \citet{EDR}:
\begin{itemize}

\item
Find all of the object's Petrosian radii, as described above.

\item
If there is no Petrosian radius, $\Pratio$ must be above $\RPcrit$ at the last
measured point in the profile (remember that $\RP(0) \equiv 1$). We
thus take $\RP$ to be the outermost measured point in the profile
(this is equivalent to assuming that the surface brightness is exactly
zero beyond this point). We then set the \texttt{NOPETRO} and
\texttt{NOPETRO\_BIG} flags and proceed to measuring other quantities.

\item Otherwise, reject all the values of $\RP$ where the corresponding
surface brightness (as estimated by differentiating the spline
representation of the cumulative surface brightness) is below
$\SBmin = 25\,\rm mag\,arcsec^{-2}$. If any values are rejected, we set the
\texttt{PETROFAINT} flag.  

\item
Keep the largest surviving $\RP$. If there is more than one,
set the \texttt{MANYPETRO} flag.

\item
If there are no surviving radii, set the \texttt{NOPETRO} flag;
set $\RP = \RPsmall = 3''$.
\end{itemize}

Once we know $\RP$, we can estimate its error $\sigma_\RP$.  We find
the Petrosian radii (following the above
prescription) corresponding to Petrosian ratios $\Pratio + \sigma_\Pratio$ and
$\Pratio - \sigma_\Pratio$.  Half the difference between them is the
estimated error on $\RP$.  
This simple approach
ignores covariances between the estimates, but gives correct errors
within 20\% as determined from repeat
measurements. 

The Petrosian flux $F_P$ is defined as the flux within $\PFluxFactor
\times \RP$; in all bands the $\RP$ used is that measured in the $r$
band. If $\PFluxFactor \times \RP$ exceeds the last measured point on
the profile, the total flux to that point is used (once again, this
corresponds to assuming that the surface brightness falls to zero at
this point, as this is the only reasonable assumption we could make). 
This happens for only 2\% of galaxies brighter than the spectroscopic limit.
The error in the Petrosian flux $\sigma_{F_P}$ is made up of
two terms, added in quadrature: The photon noise within $\PFluxFactor
\RP$ due to the object and sky, and a term due to the uncertainty in
$\RP$. This second term is $0.5\left[C(\RP + \sigma_\RP) - C(\RP -
\sigma_\RP)\right]$, where $C$ is the cumulative profile as above.  We
neglect the covariance between these two terms; the contribution to
the photon noise from the region between $\RP + \sigma_\RP$ and $\RP -
\sigma_\RP$ is negligible; the uncertainty in $\RP$ is also mostly
determined locally, and is thus more-or-less uncorrelated with the
Poisson term.  Both terms are included in all bands, even though the
Petrosian aperture is based on the $r$ band Petrosian radius for all
bands. \S~\ref{sec:reproducibility} showed that the resulting errors
are quite accurate. 

We also calculate two concentration parameters $\R_{50}$ and $\R_{90}$,
the radii containing 50\% and 90\% of the Petrosian flux. Their errors
are na\"\i{}vely estimated as e.g.  $0.5(\R_{50,F_P + \sigma_{F_P}} -
\R_{50,F_P - \sigma_{F_P}})$ where $\R_{50,F_P + \sigma_{F_P}}$ is the
value of $\R_{50}$ that we'd estimate if the Petrosian flux were $F_P +
\sigma_{F_P}$.  Repeat measurements show that the errors in $\R_{50}$
are overestimated by a factor of two at $r^*=18$, while the errors in
$\R_{90}$ are correct to 10\%.

\section{On the fraction of galaxy targets crossing the magnitude
limits in repeated scans}
\label{sec:overlap-integral}

The Petrosian magnitudes have finite errors, and thus a sharp cut in
{\em observed} magnitude will be a slightly fuzzy cut in true
magnitudes.  One effect, as we saw in \S~\ref{sec:reproducibility}, is
that samples defined from repeat imaging scans of the same area of sky
will not be identical.  We quantify the expected effect here. 

The probability that a galaxy with a true magnitude $m$ is observed to be
brighter than the magnitude limit $m_{l}$ in one scan and fainter than the
magnitude limit in another scan is given by
\be
P(m) = p(m)\left[1-p(m)\right],
\ee
where 
\be
p(m) = \frac{1}{\sqrt{2\pi}\sigma_m}\int_{-\infty}^{m_l}
e^{-\frac{\left(m-m_l\right)^{2}}{2\sigma_m^2}}dm_l \equiv
\frac{1}{2}{\rm erfc}\left(\frac{m-m_l}{\sqrt{2}\sigma_m}\right)
\ee
is the probability that a galaxy is brighter than the magnitude limit in
one scan, and $\sigma_m$ is the error in photometry (in magnitudes), which we
assume is distributed as a Gaussian.
Hence, the fraction of galaxy targets that are targeted in one scan
but not in the other is
\be
F (m_1 < m_l, m_2 > m_l) = \frac{\int_{-\infty}^{\infty} n(m)p(m)\left[1-p(m)\right]}{\int_{-\infty}^{\infty} n(m)p(m)},
\label{eq:scatter}
\ee
where $n(m)$ is the differential number counts of galaxies as a function
of magnitude.

Yasuda \etal\ (2001) have found that $n(m) \propto 10^{0.55m}$ near
the magnitude limit of $m_l = 17.77$ in the $r$-band.  At this
magnitude, $\sigma_m = 0.035$ mag, and equation~(\ref{eq:scatter})
predicts that about $3.2\%$ of galaxies to cross the magnitude
limit in two repeated scans, due to random photometric errors alone.
This predicted fraction is very close to the fraction 3.7\% found
in the test discussed in \S\ref{sec:reproducibility}.

\begin{figure}
\plotone{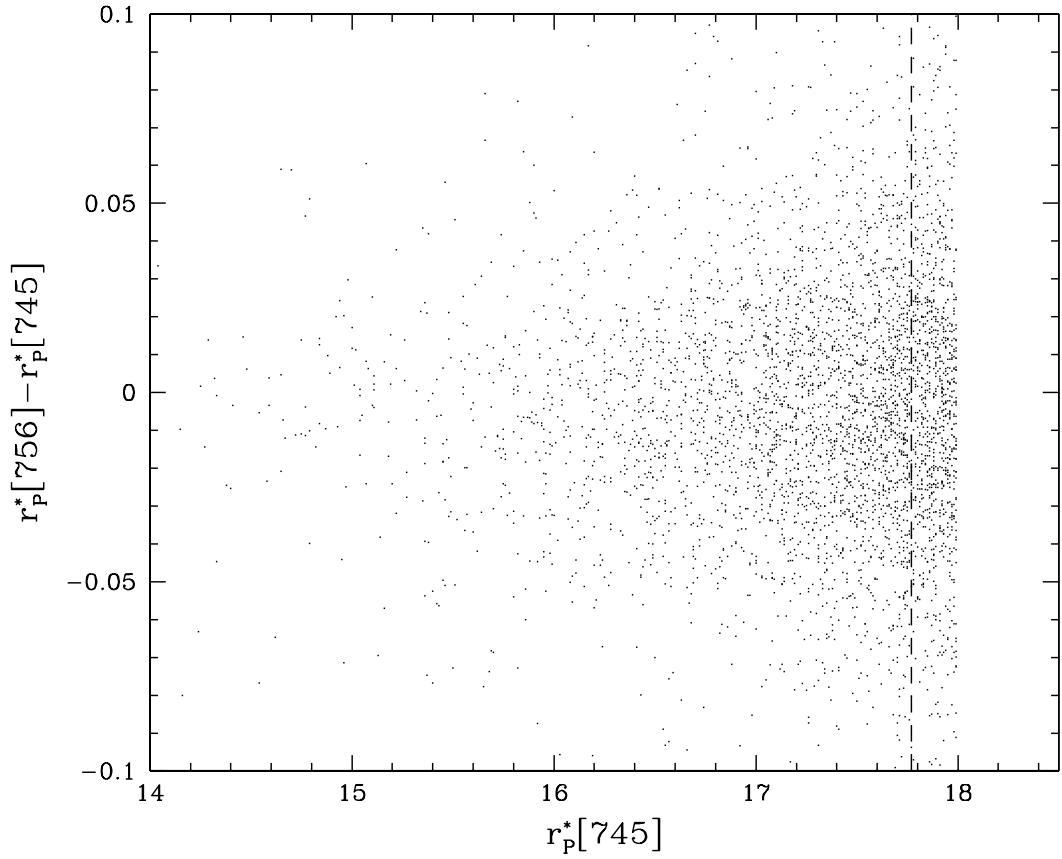}
\caption{The difference between the $r$-band Petrosian magnitudes 
of the same galaxies measured in two different imaging runs.
The dashed line shows the magnitude limit $r_P = 17.77$  of the
main galaxy sample. 
}
\label{fig:magerr}
\end{figure}

\end{document}